\documentclass[Afour, sageh, times]{sagej}

\usepackage{moreverb,url}
\usepackage[colorlinks,bookmarksopen,bookmarksnumbered,citecolor=red,urlcolor=red]{hyperref}

\def\journalname{The International Journal of High Performance Computing Applications}
\def\volumeyear{2020}

\usepackage[utf8]{inputenc}
\usepackage{multirow}
\usepackage{amsmath, amssymb, amsfonts, commath}

\usepackage{algorithmic}
\usepackage[ruled, linesnumbered]{algorithm2e}
\SetAlFnt{\small}
\SetAlCapFnt{\small}
\SetAlCapNameFnt{\small}
\DontPrintSemicolon 

\usepackage{graphicx}
\usepackage[font=small, justification=centering]{caption}
\usepackage[caption=false, justification=centering]{subfig}
\usepackage{textcomp}
\usepackage{xcolor}
\definecolor{lssblue}{rgb}{0.121,0.231,0.4}
\definecolor{lssred}{rgb}{0.76,0.3,0.19}

\usepackage{booktabs}
\usepackage{rotating}
\usepackage{color, colortbl}
\usepackage[colorinlistoftodos, textwidth=3cm, textsize=footnotesize]{todonotes}
\usepackage{a4wide}
\usepackage{nicefrac}
\usepackage{siunitx}

\usepackage{minted} 
\input{listings/output/pygmentsDefs}
\expandafter\def\csname PY@tok@err\endcsname{}

\usepackage[noabbrev]{cleveref}
\crefname{algocf}{alg.}{algs.}

\usepackage{pgfplots}
\usetikzlibrary{decorations.pathreplacing, angles, quotes, arrows, positioning, backgrounds, shapes}
\pgfplotsset{compat=newest}
\usepgfplotslibrary{groupplots}
\usepgfplotslibrary{dateplot}
\usetikzlibrary{patterns}

\tikzset{font={\small}}

\pgfdeclarelayer{background}
\pgfdeclarelayer{foreground}
\pgfsetlayers{background,foreground}

\definecolor{mybluei}{RGB}{124,156,205} 
\definecolor{myblueii}{RGB}{73,121,193}
\definecolor{mygreen}{RGB}{202,217,126}

\definecolor{color1}{HTML}{5A88AC} 
\definecolor{color2}{HTML}{75B1B0} 
\definecolor{color3}{HTML}{C5B353} 
\definecolor{color4}{HTML}{D18855} 

\pgfdeclarelayer{background}
\pgfsetlayers{background,main}

\newcommand{\todURi}[1][]{\todo[color=green!20, author=Uli, size=\footnotesize, inline]}
\newcommand{\todUR}[1][]{\todo[color=green!20, size=\footnotesize, linecolor=darkgray, author=Uli]}

\newcommand{\lbmpy}{{{\em lbmpy}}}

\newcommand{\walberla}{\textsc{waLBerla}}
\newcommand{\nvidia}{NVIDIA}
\newcommand{\pizdaint}{Piz~Daint}

\setcitestyle{aysep={,}}

\begin{document}

\runninghead{Holzer et al.}

\title{Highly Efficient Lattice-Boltzmann Multiphase Simulations 
	of Immiscible Fluids at High-Density Ratios on CPUs and GPUs through Code Generation}

\author{Markus Holzer\affilnum{1}, Martin Bauer\affilnum{1}, Harald Köstler\affilnum{1} and Ulrich Rüde\affilnum{1}$^,$\affilnum{2}}

\affiliation{\affilnum{1}Chair for System Simulation, Friedrich–Alexander–Universit\"at Erlangen–N\"urnberg, Cauerstra\ss{}e 11, 91058 Erlangen, Germany\\
\affilnum{2}CERFACS, 31057 Toulouse Cedex 1, France}

\corrauth{Markus Holzer, Chair for System Simulation, Friedrich–Alexander–Universit\"at Erlangen–N\"urnberg, Cauerstra\ss{}e 11, 91058 Erlangen, Germany}

\email{markus.holzer@fau.de}

\begin{abstract}
A high-performance implementation of a multiphase lattice Boltzmann method 
based on the conservative Allen-Cahn model 
supporting high-density ratios and high Reynolds numbers is presented. 
Metaprogramming techniques are used to generate optimized code for CPUs and GPUs automatically. 
The coupled model is specified in a high-level symbolic description and optimized through automatic transformations. 
The memory footprint of the resulting algorithm is reduced through the fusion of compute kernels.
A roofline analysis demonstrates the excellent efficiency of the generated code on a single GPU. 
The resulting single GPU code 
has been integrated into the multiphysics framework \walberla{} to run massively parallel simulations on 
large domains. 
Communication hiding and GPUDirect-enabled MPI yield near-perfect scaling behaviour. 
Scaling experiments are conducted on the \pizdaint{} supercomputer with up to $2048$~GPUs, 
simulating several hundred fully resolved bubbles. 
Further, validation of the implementation is shown in a physically relevant 
scenario--a three-dimensional rising air bubble in water. 
\end{abstract}


\maketitle

\section{Introduction}
The numerical simulation of multiphase flow 
is a challenging field of computational fluid dynamics \citep[see][]{prosperetti_tryggvason_2007}.
Although 
a wide variety of different approaches have been developed, 
simulating the dynamics of immiscible fluids with high-density ratio and at high Reynolds numbers is still considered complicated \citep[see][]{multiphase_LB_book}. 
Such multiphase flows 
require 
models for the 
interfacial dynamics \citep[see][]{YAN2011649}.
A full resolution of 
these phenomena is usually impractical 
for macroscopic CFD techniques
since the interface is only a few nanometers thick as pointed out by \citet{PhysRevE.96.053301}.

Therefore, sharp interface techniques model the interface as two-sided boundary conditions
on a free surface and can thus 
achieve a discontinuous transition \citep[see][]{thuerey, Pohl, BOGNER20151}.
The modelling and implementation of these boundary conditions can be complicated as stated by \citet{Bogner2016}, 
especially in a parallel setting.
Diffuse-interface models, in contrast, represent the interface in a transition region
of a finite thickness that is typically much wider than the true physical interface \citep[see][]{Anderson98}. 
Thus, the sharp interface between the fluids is replaced by a smooth transition
with thickness $\xi$ of a few grid cells. 
This removes abrupt jumps and certain singularities coming along with a sharp interface. 
In this work, we will focus on phase-field modeling. 
Here, an advection-diffusion-type equation is solved based on either 
Cahn-Hilliard \citep[see][]{Hilliard58} or Allen-Cahn models (ACM) \citep[see][]{ALLEN1976425} to track the interface. 
The governing equations can 
be solved using the lattice Boltzmann method (LBM)
that is based on the kinetic theory expressed by the original Boltzmann equation \citep[see][]{LI2016}.

Typically, the simulation domain is discretized with a Cartesian grid for 3D LBM simulations. 
Due to its explicit time-stepping scheme and its high data locality, the LBM is well suited for extensive parallelization \citep[see][]{waLBerla}.
For the simulation of multiphase flows, additional force terms are needed 
that employ non local data points so that the ideal situation may worsen somewhat.
Fakhari et al.~have shown how to improve the locality for the conservative 
ACM \citep[see][]{PhysRevE.96.053301}. 
In order to resolve physically relevant phenomena, a sufficiently high resolution at the
interface between the fluid phases is necessary.
Thus we may need simulation domains containing billions of  grid points and beyond. 
This creates the need for using high-performance computing (HPC) systems and for a highly parallelized and run-time efficient algorithms.
However, highly-optimized implementations in low-level languages 
like Fortran or C often suffer from poor flexibility and poor extensibility
so that algorithmic changes and the development of new models may get tedious. 

To overcome this problem, we here use a work flow where we
generate the compute intensive kernels at compile time with a 
code generator realized in Python.
Thus we obtain the highest possible performance while maintaining maximal flexibility \citep[see][]{Bauer19}. 
Furthermore, the symbolic description of the complete ACM allows us
to use \emph{IPython} \citep[see][]{IPython} as an interactive prototyping environment. 
Thus, changes in the model, like additional terms, different discretization schemes, different versions of
the LBM or different LB stencils can be incorporated directly on the level of the defining mathematical equations.
These equations can be represented in \LaTeX-form.
The generated code can then
run in parallel with OpenMP parallel on a single CPU or on a single GPU. 
Once a working prototype has created in this working mode, 
the automatically generated code kernels can be integrated into existing HPC 
software as external C++ files. 
In this way, we can  execute massively parallel simulations.
Note that this workflow permits to describe physical models in symbolic form
and still run with maximal efficiency on parallel supercomputers,
Using code generation, we realize a higher level ob abstraction and thus an improved separation of concerns.

The remainder of the article is structured  as follows. 
In \cref{sec:related_work}, we will summarize related work to the conservative ACM. 
In \cref{sec:model_description}, we will introduce the governing equations of the conservative ACM. \Cref{sec:flexibleImpl} presents details of the implementation, by first introducing the code generation toolkit \lbmpy{} by \citet{bauer2020lbmpy} which constitutes the basis of our implementation. 
Then, we present the phase-field algorithm itself in a straightforward and an improved form, where the improvements essentially lie in the minimization of the memory footprint. 
This is primarily achieved by changing the structure of the algorithm in order to be able to 
fuse several compute kernels. 
The performance of our implementation is discussed in \cref{sec:benchmark}. 
We first show a comparison of the straightforward and the improved algorithm. 
Then the performance of the improved version is analyzed on a single GPU with a roofline approach. 
After that the scaling behavior on up to $2048$~GPU nodes is presented in \cref{sec:weak_scaling}. 
For the scaling on many GPU nodes, communication hiding strategies are explained and analyzed. 
Finally, we validate the physical correctness of our implementation 
with test cases for a rising bubble in \cref{sec:single_rising_bubble,sec:bubble_field}. 

\section{Related Work}
\label{sec:related_work}
The interface tracking in this work is carried out with the Allen-Cahn equation (ACE) \citep[see][]{ALLEN1976425}. A modification of the ACE to a phase-field model was proposed by \citet{SUN2007626}.
Nevertheless, it is Cahn-Hilliard theory \citep[see][]{Hilliard58}, which is most often used in phase-field models to perform the interface tracking. 
A reason for that is the implicit conservation of the phase-field and thus the conservation of mass. 
As a drawback, it includes fourth-order spatial derivatives, which worsens the locality of the LB framework as pointed out by \citet{PhysRevE.91.063309}. 
In order to make the ACE accessible for phase-field models, \citet{CHIU2011185}~have presented it in conservative form. 
Furthermore, the conservative ACE contains only second-order derivatives which allows a more efficient implementation. \par 

In the work of \citet{PhysRevE.91.063309} the conservative ACE was first solved using a single relaxation time (SRT) algorithm. 
Additionally, they proposed an improvement of the algorithm by solving the collision step in the central moment space and adapting the equilibrium formulation which makes it possible to directly calculate the gradient of the phase-field locally via the moments. 
This promising approach, however, leads to a loss of accuracy \citep[see][]{FAKHARI20191154}. 
On the other hand \citet{PhysRevE.96.053301} used an SRT formulation to solve the conservative ACE with isotropic finite-differences \citep[see][]{Kumar2004} to compute the curvature of the phase-field. 
This approach was later extended by \citet{MITCHELL20181} for the three-dimensional case. 
A disadvantage of this approach is, that it becomes complicated to apply single array streaming patterns \citep[see][]{Wittmann} like the AA-pattern or the Esoteric Twist to the LBM as stated by \citet{PhysRevE.91.063309}. 
This is due to a newly introduced non-locality in the update process, originating in the finite difference calculation \citep[see][]{PhysRevE.91.063309}. 
In this publication, the phase-field LBE is presented with a multiple relaxation time (MRT) formulation which was first published by \citet{PhysRevE.94.023311}. 
This formulation was also used in recent studies by \citet{doi:10.1063/1.5100215}.

\section{Model Description}
\label{sec:model_description}
\subsection{LB Model for Interface Tracking}
As described by \citet{PhysRevE.96.053301}, 
the phase-field $\phi$ in the conservative ACM assumes two extreme values, 
	$\phi_L$ and $\phi_H$,
in the bulk of the lighter and the heavier fluid.
There are different possibilities to choose these values \citep[see][]{PhysRevE.96.053301}.
Throughout the simulations in this work we set $\phi_L = 0$ and $\phi_H = 1$, respectively.
The phase-field equation for two immiscible fluids reads
\begin{align}
\label{eq:allen_cahn}
\frac{\partial \phi}{\partial t} + \nabla \cdot \phi \boldsymbol{u} = \nabla \cdot M \left[\nabla \phi - \frac{\nabla \phi}{\abs{\nabla \phi}} \theta \right],
\end{align}
where $\theta = \nicefrac{1 - 4 \left(\phi - \phi_0\right)^2}{\xi}$, $t$ is the time, $\boldsymbol{u}$ is the macroscopic velocity, $\xi$ is the interface thickness and $\phi_0 = \nicefrac{(\phi_L + \phi_H)}{2}$ indicates the location of the phase-field. Further, the mobility is
\begin{align}
\label{eq:mobility}
M = \tau_\phi c_s^2 \Delta t.
\end{align}
is related to the phase-field relaxation time $\tau_\phi$. The speed of sound $c_s = \nicefrac{c}{\sqrt{3}}$, where $c = \nicefrac{\Delta x}{\Delta t}$, and $\Delta x = \Delta t = 1$, which is common practice for uniform grids \citep[see][]{PhysRevE.96.053301}.

In the equilibrium state the profile of the phase-field $\phi$ of an interface located at $\boldsymbol{x_0}$ is 
\begin{align}
\phi (\boldsymbol{x}) = \phi_0 \pm \frac{\phi_H - \phi_L}{2} \tanh{\left(\frac{\boldsymbol{x} - \boldsymbol{x_0}}{\nicefrac{\xi}{2}}\right)}.
\end{align}
The LB model for \cref{eq:allen_cahn} to update the phase-field distribution function $h_i$ can be written as \citep[see][]{PhysRevE.91.063309}
\begin{align}
\label{eq:lb_interface}
h_i (\boldsymbol{x} + \boldsymbol{e}_i \Delta t, t + \Delta t) = & \Omega_{ij}^h(h_j^{\mathrm{eq}} - h_j - \frac{1}{2} F_j^\phi)|_{(\boldsymbol{x}, t)} \nonumber \\ 
& +h_i(\boldsymbol{x}, t) + F_i^\phi(\boldsymbol{x}, t),
\end{align}
in which the forcing term is given by \citep[see][]{PhysRevE.96.053301}
\begin{align}
\label{eq:interfaceTracking}
F_i^\phi (\boldsymbol{x}, t) = \Delta t \, \theta \, w_i \, \boldsymbol{e}_i \cdot \frac{\nabla \phi}{\abs{\nabla \phi}}.
\end{align}
In \cref{eq:lb_interface} $\Omega_{ij}^h$ represents the elements of the
collision matrix and takes the form 
	$\boldsymbol{\Omega} = \boldsymbol{M}^{-1} \boldsymbol{S} \boldsymbol{M}$,
where
$\boldsymbol{M}$ is the moment matrix \citep[see][]{PhysRevE.87.023304} and
$\boldsymbol{S}$ is the diagonal relaxation matrix.
As described by \citet{PhysRevE.94.023311} we relax the first order moments by
	$\nicefrac{1}{\tau_\phi}$
and all other moments by one.
The phase-field relaxation time $\tau_\phi$ is calculated with \cref{eq:mobility}. 
Further, the speed of sound
	$c_s = \nicefrac{c}{\sqrt{3}}$,
where $c = \nicefrac{\Delta x}{\Delta t}$, and $\Delta x = \Delta t = 1$, which is common practice for uniform grids \citep[see][]{PhysRevE.96.053301}. 
The parameters $w_i$ and $\boldsymbol{e}_i$ correspond to the lattice weights and the mesoscopic velocity.
The equilibrium phase-field distribution function $h_i^{\mathrm{eq}} = \phi \Gamma_i$, where
\begin{align}
	\Gamma_i = w_i \left[1 + \frac{\boldsymbol{e}_i \cdot \boldsymbol{u}}{c_s^2} +
		\frac{\left(\boldsymbol{e}_i \cdot \boldsymbol{u}\right)^2}{2 c_s^4} -
		\frac{\boldsymbol{u} \cdot \boldsymbol{u}}{2 c_s^2}\right]
\end{align}
is the dimensionless distribution function.
By taking the zeroth moment of the phase-field distribution functions the phase-field $\phi$ can be evaluated
\begin{align}
\label{eq:updatePhaseField}
	\phi = \sum_{i} h_i.
\end{align}
The density $\rho$ for the whole domain is calculated by using a linear interpolation
\begin{align}
\label{eq:density}
\rho &= \rho_L + (\phi - \phi_L)(\rho_H - \rho_L),
\end{align}
where $\rho_H$ and $\nu_H$ are the density and the kinematic viscosity of the heavier fluid
while $\rho_L$ and $\nu_L$ are the density and the kinematic viscosity of the lighter fluid.
\subsection{LB Model for Hydrodynamics}
\label{sec:LB_hydro}
In a macroscopic form the continuity and the incompressible Navier-Stokes
equations describe the evolution of a flow field and can be written as 
\begin{equation}
\begin{aligned}
\label{eq:NSE}
\nabla \cdot \boldsymbol{u} &= 0 \\
\rho \left(\frac{\partial \boldsymbol{u}}{\partial t} + \boldsymbol{u} \cdot \nabla \boldsymbol{u}\right) &= - \nabla p + \nabla \cdot \Pi + \boldsymbol{F},
\end{aligned}
\end{equation}
where $\rho$ is the density, $p$ the pressure,
	$\Pi = \mu \left(\nabla \boldsymbol{u} + \nabla \boldsymbol{u}^T\right)$
the viscous stress tensor, $\mu$ the dynamic viscosity and 
	$\boldsymbol{F} = \boldsymbol{F}_s + \boldsymbol{F}_b$
are the surface tension force and external forces respectively.
To solve \cref{eq:NSE} the following LB model is used \citep[see][]{doi:10.1063/1.5100215}
\begin{align}
\label{eq:lb_flow}
g_i (\boldsymbol{x} + \boldsymbol{e}_i \Delta t, t + \Delta t) = & \Omega_{ij}^g(g_j^{\mathrm{eq}} - g_j - \frac{1}{2} F_j)|_{(\boldsymbol{x}, t)}  \nonumber \\ 
& + g_i(\boldsymbol{x}, t) + F_i(\boldsymbol{x}, t),
\end{align}
where the hydrodynamic forcing is given by
\begin{align}
F_i (\boldsymbol{x}, t) = \Delta t w_i \frac{\boldsymbol{e}_i \boldsymbol{F}}{\rho c_s^2},
\end{align}
and $g_i$ is the velocity-based distribution function for incompressible fluids.
The equilibrium distribution function is 
\begin{align}
g_i^{\mathrm{eq}} = p^* w_i + \left(\Gamma_i - w_i\right),
\end{align}
where $p^* = \nicefrac{p}{\rho c_s^2}$ donates the normalized pressure. The hydrodynamic force $\boldsymbol{F}$ consists of four terms 
\begin{align}
\label{eq:forceHydro}
\boldsymbol{F} = \boldsymbol{F}_p + \boldsymbol{F}_s + \boldsymbol{F}_\mu + \boldsymbol{F}_b.
\end{align}
The pressure force can be obtained as
\begin{align}
\label{eq:pressure_force}
\boldsymbol{F}_p = -p^* c_s^2 \nabla \rho,
\end{align}
where the normalized pressure is calculated as the zeroth moment of the hydrodynamic distribution function 
\begin{align}
p^* = \sum_{i} g_i.
\end{align}
The surface tension force 
\begin{align}
\label{eq:surfaceTension}
\boldsymbol{F}_s = \mu_\phi \nabla \phi,
\end{align}
is the product of the chemical potential 
\begin{align}
\label{eq:chemPot}
\mu_\phi = 4 \beta (\phi - \phi_L) (\phi - \phi_H) (\phi - \phi_0) - \kappa \nabla^2 \phi
\end{align}
and the gradient of the phase-field. The coefficients $\beta = \nicefrac{12 \sigma}{\xi}$ and $\kappa = \nicefrac{3 \sigma \xi}{2}$ link the interface thickness and the surface tension. 
For an MRT scheme the viscous force is computed as
\begin{align}
\label{eq:viscousForce}
\begin{split}
F_{\mu, i}^{\mathrm{MRT}} = &- \frac{\nu}{c_s^2 \Delta t} \left[\vphantom{\sum} \sum_{\alpha} e_{\alpha i} e_{\alpha j} \right.\\
&\left. \times \sum_{\beta} \Omega_{\alpha \beta}(g_\beta - g_\beta^{\mathrm{eq}}) \vphantom{\sum}\right] \frac{\partial \rho}{\partial x_j},
\end{split}
\end{align}
where the viscosity $\nu$ is related to hydrodynamic relaxation time $\tau$
\begin{align}
\nu = \tau c_s^2 \Delta t.
\end{align}
There are a few different ways to interpolate the hydrodynamic relaxation time as
it is shown in the work of \citet{PhysRevE.96.053301}. 
Overall, they have demonstrated to get the most stable results with a linear interpolation.
Therefore, we will use it in this work
\begin{align}
\tau = \tau_L + (\phi - \phi_L)(\tau_H - \tau_L).
\end{align}
We relax the second order moments with the hydrodynamic relaxation rate
\begin{align}
s_\nu = \frac{1}{\tau + \nicefrac{1}{2}},
\end{align}
when solving \cref{eq:lb_flow} to ensure the correct viscosity of the fluid.
All other moments are relaxed by one.
The velocity $\boldsymbol{u}$ is obtained via the first moments of
the hydrodynamic distribution function and gets shifted by the external forces
\begin{align}
\label{eq:updateVelocityField}
\boldsymbol{u} = \sum_{i} g_i \boldsymbol{e}_i + \frac{\boldsymbol{F}}{2 \rho} \Delta t.
\end{align}

In order to approximate the gradient in \cref{eq:interfaceTracking,eq:pressure_force,eq:surfaceTension,eq:viscousForce} a second order isotropic stencil can be applied \citep[see][]{Kumar2004, Ramadugu_2013}
\begin{align}
\label{eq:grad}
\nabla \phi = \frac{c}{c_s^2 \left(\Delta x\right)^2} \sum_{i} \boldsymbol{e}_i w_i \phi \left(\boldsymbol{x} + \boldsymbol{e}_i \Delta t, t\right).
\end{align}
The Laplacian in \cref{eq:chemPot} can be approximated with
\begin{align}
\label{eq:lap}
\nabla^2 \phi = \frac{2 c^2}{c_s^2 \left(\Delta x\right)^2} \sum_{i} w_i \left[\phi \left(\boldsymbol{x} + \boldsymbol{e}_i \Delta t, t\right) - \phi \left(\boldsymbol{x}, t\right)\right].
\end{align}

\section{Software Design for a Flexible Implementation}
\label{sec:flexibleImpl}

\subsection{Code Generation}
\label{sec:codeGeneration}

Our implementation of the conservative ACM is based on the open source LBM code generation
framework \lbmpy{}\footnote{https://i10git.cs.fau.de/pycodegen/lbmpy} \citep[see][]{bauer2020lbmpy}.
Using this meta-programming approach, we address the often encountered trade-off between code flexiblity, readability, 
and maintainability on the one hand, and platform-specific performance engineering on the other hand. 
Especially when targeting modern heterogeneous HPC architectures,
a highly optimized compute kernel, 
may require that loops are unrolled, common subexpressions extracted, and possibly
hardware-specific intrinsics are used.
In state-of-the art optimized software \citep[see][]{HagerAndWellein}, these transformations are essential,
and must be performed manually for each target architecture.
Clearly, the resulting codes are time-consuming to develop, are error prone, hard to read, difficult maintain
and often very hard to adapt and extend.
Flexibility and maintainability have been sacrificed, since
such complex programming techniques are essential to get the full performance available on the system. 

Here, in contrast, we employ 
the LBM code generation framework \lbmpy{}.
Thanks to the automated code transformations, the LB scheme 
can be specified in a high-level symbolic representation.
The hardware- and problem-specific transformations are applied automatically
so that starting form an abstract representation,
highly efficient C code for CPUs or CUDA/OpenCL code for GPUs can be generated with little effort.

Our new tool \lbmpy{} is realized as a Python package that in turn is 
built by using the 
stencil code generation and transformataion framework 
~\emph{pystencils}\footnote{https://github.com/mabau/pystencils/} \citep[see][]{Bauer19}.
The flexibility of \lbmpy{} results from the fully symbolic representation of collision operators and compute kernels, utilizing the computer algebra system \emph{SymPy}\citep[see][]{sympy}.
The package offers an interactive environment for method prototyping and 
development on a single workstation, similar to what
FEniCS \citep[see][]{AlnaesBlechta2015a} is in the context of finite element methods.
Generated kernels can then be easily integrated into the HPC framework \walberla{}, 
which is designed to run massively parallel simulations for a wide range of scientific applications \citep[see][]{waLBerla}.
In this workflow, \lbmpy{} is employed for generating optimized compute and communication kernels,
whereas \walberla{} provides the software structure to use these kernels in large scale scenarios on supercomputers. `\lbmpy{} can generate kernels for moment-based LB schemes,
namely single-relaxation-time (SRT), two-relaxation-times (TRT),
and multiple-relaxation-time (MRT) methods. 
Additionally, modern cumulant and entropically stabilized collision operators are 
supported.

\begin{figure}[htb]
	\centering
	\resizebox{\linewidth}{!}
	{
		\begin{tikzpicture}[framed,background rectangle/.style={fill=lssblue!10, rounded corners, draw=black!90, dashed}]

\def\scale{1.4};

\def\wi{1};

\def\lw{.75mm};

	\tikzset{
		state/.style={
			rectangle split,
			rectangle split parts=2,
			rounded corners,
			draw=black, very thick,
			minimum height=2em,
			text width=3.3cm,
			inner sep=2pt,
			text centered,
		}
	}

	\node (a) [state, rectangle split part fill={color1,gray!20}] at (2,10) {\textbf{Stencils}  \nodepart[align=right]{two} 
		 D2Q9 \textcolor{color1}{\scalebox{\scale}{$\bullet$}}\\
		 D3Q15 \textcolor{color1}{\scalebox{\scale}{$\bullet$}}\\
		 D3Q19 \textcolor{color1}{\scalebox{\scale}{$\bullet$}}\\
		 D3Q27 \textcolor{color1}{\scalebox{\scale}{$\bullet$}}};
	 
	\node (b) [state, below = .5cm of a, rectangle split part fill={color2,gray!20}] {\textbf{Collision methods}  \nodepart[align=right]{two} 
		SRT \textcolor{color2}{\scalebox{\scale}{$\bullet$}}\\
		TRT \textcolor{color2}{\scalebox{\scale}{$\bullet$}}\\
		MRT \textcolor{color2}{\scalebox{\scale}{$\bullet$}}\\
		Cumulant \textcolor{color2}{\scalebox{\scale}{$\bullet$}} \\
	    Entropic KBC \textcolor{color2}{\scalebox{\scale}{$\bullet$}}};
	
	\node (c) [state, below = .5cm of b, rectangle split part fill={color3,gray!20}] {\textbf{Streaming pattern}  \nodepart[align=right]{two} 
		Collide only \textcolor{color3}{\scalebox{\scale}{$\bullet$}}\\
		Stream pull collide \textcolor{color3}{\scalebox{\scale}{$\bullet$}}\\
		Collide stream push \textcolor{color3}{\scalebox{\scale}{$\bullet$}}\\
		Esoteric Twist \textcolor{color3}{\scalebox{\scale}{$\bullet$}} \\
	    AA-pattern \textcolor{color3}{\scalebox{\scale}{$\bullet$}}};

	\node (d) [state, below = .5cm of c, rectangle split part fill={color4,gray!20}]  {\textbf{Force model} \nodepart[align=right]{two} 
		Custom model 1 \textcolor{color4}{\scalebox{\scale}{$\bullet$}}\\
		Custom model 2 \textcolor{color4}{\scalebox{\scale}{$\bullet$}}};
	
	\node [state, below right=-2cm and \wi of a, rectangle split part fill={lssblue!30,gray!20}] {\textbf{Phase-field LB} \nodepart[align=left]{two} 
		\textcolor{color1}{\scalebox{\scale}{$\bullet$}} Stencil \\
		\textcolor{color2}{\scalebox{\scale}{$\bullet$}} Collision method \\
		\textcolor{color3}{\scalebox{\scale}{$\bullet$}} Streaming pattern \\
		\textcolor{black}{\scalebox{\scale}{$\bullet$}} Raw moments\\
		\textcolor{black}{\scalebox{\scale}{$\bullet$}} Compressible \\
		\textcolor{black}{\scalebox{\scale}{$\bullet$}} Relaxation time $\omega_\phi$ \\
		\textcolor{black}{\scalebox{\scale}{$\bullet$}} Discrete equilibrium \\
		\textcolor{color4}{\scalebox{\scale}{$\bullet$}} Force model};

	\node [state,above right=-.9cm and \wi of d, rectangle split part fill={lssblue!30,gray!20}] {\textbf{Hydrodynamic LB}  \nodepart[align=left]{two} 
		\textcolor{color1}{\scalebox{\scale}{$\bullet$}} Stencil \\
		\textcolor{color2}{\scalebox{\scale}{$\bullet$}} Collision method \\
		\textcolor{color3}{\scalebox{\scale}{$\bullet$}} Streaming pattern \\
		\textcolor{black}{\scalebox{\scale}{$\bullet$}} Weighted moments\\
		\textcolor{black}{\scalebox{\scale}{$\bullet$}} Incompressible \\
		\textcolor{black}{\scalebox{\scale}{$\bullet$}} Relaxation matrix $\boldsymbol{S}$ \\
		\textcolor{black}{\scalebox{\scale}{$\bullet$}} Discrete equilibrium \\
		\textcolor{color4}{\scalebox{\scale}{$\bullet$}} Force model};
	
\draw[-stealth, line width= \lw, color1]  (3.742, 10.04) to[out=0, in=180]  (3.742+\wi, 10.35);
\draw[-stealth, line width= \lw, color2]  (3.742, 7.86) to[out=0, in=180]  (3.742+\wi, 9.97);
\draw[-stealth, line width= \lw, color3]  (3.742, 4.45) to[out=0, in=180]  (3.742+\wi, 9.6);
\draw[-stealth, line width= \lw, color4]  (3.742, 1.9) to[out=0, in=180]  (3.742+\wi, 7.65);

\draw[-stealth, line width= \lw, color1]  (3.742, 9.27) to[out=0, in=180]  (3.742+\wi, 4.55);
\draw[-stealth, line width= \lw, color2]  (3.742, 7.1) to[out=0, in=180]  (3.742+\wi, 4.15);
\draw[-stealth, line width= \lw, color3]  (3.742, 4.08) to[out=0, in=180]  (3.742+\wi, 3.79);
\draw[-stealth, line width= \lw, color4]  (3.742, 1.5) to[out=0, in=180]  (3.742+\wi, 1.84);

\end{tikzpicture}
	}
	\caption{Flexibility of the conservative ACM with the \lbmpy{} code generation framework. The boxes on the right show the two LB steps. On the left options are shown which can be applied to the two LB steps by \lbmpy{}. The connecting lines show a possible configuration which will be used for the benchmark in this section.}
	\label{fig:lbmpystructure}
\end{figure}
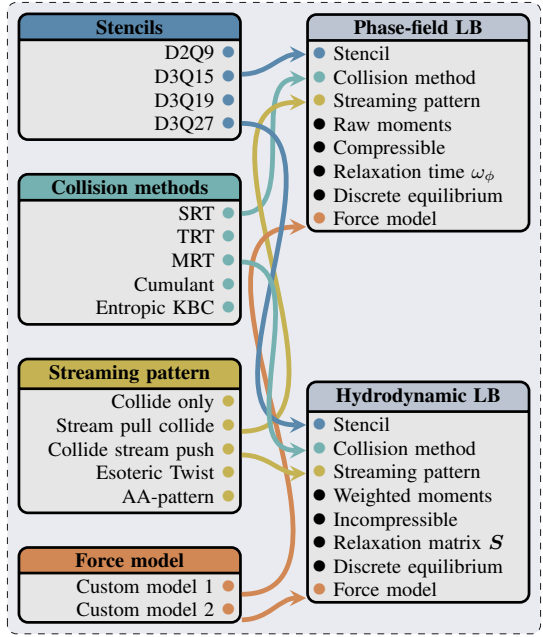

When implementing the coupled multiphase scheme as described in \cref{sec:model_description} with \lbmpy{}, 
we can reuse several major building blocks that are already part of \lbmpy{} (see \cref{fig:lbmpystructure}). 
First we can choose between different single-phase collision operators to use for
the Allen-Cahn and the hydrodynamic LBM. 
We can easily switch between different lattices, allowing us to quickly explore the accuracy-performance trade-off between stencils with 
more or less neighbors. 
A native 2D implementation is also quickly available by selecting the D2Q9 lattice model. 

Then, the selected collision operators of \lbmpy{} can be adapted to the specific requirements of the scheme.
In our case, we have to add the forcing terms \cref{eq:interfaceTracking,eq:forceHydro}.
This is done on the symbolic level, such that no additional arrays for storing these terms have to 
be introduced, as would typically be the case when extending an existing LB method implemented in C/C++. 
Also no additional iteration passes are needed %
to compute the force terms. 
The additional forces computed directly and are 
within the loops that update the 
LB distributions, thus significantly saving memory and operational overhead. 
Furthermore, optimization passes, like common subexpression elimination, 
SIMD vectorization via intrinsics, or CUDA index mapping are done 
automatically by transformations further down the pipeline,
the new force terms are fully included in the optimization.
Note how this leads to a clean separation of concerns between model development
and optimization with obvious benefits for code maintainability and flexibility without sacrificing
the possibility to achieve best possible performance.

On the modeling level, this code generation approach and our tools allow
the application developer 
to express the methods using 
a concise mathematical notation. 
LB collision operators are formulated in so-called collision space spanned by moments or cumulants \citep[see][]{comparison_collision_operator}. 
For each moment/cumulant a relaxation rate and its respective equilibrium value is chosen. 
For a detailed description of this formalism and its realization in Python see \citet{bauer2020lbmpy}.

Similarly, our system 
supports the mathematical formulation of differential operators
that can be discretized automatically with various numerical approximations of derivatives. 
This functionality is employed 
to express the forcing terms.
The Python formulation directly mimics the mathematical definition as shown in \cref{eq:grad,eq:lap},
i.e.~it provides a gradient and Laplacian operator. 
The user then can select between different finite difference discretizations, selecting stencil neighborhood, approximation order, and isotropy requirements.




Starting from the symbolic representations, we create the compute kernels for our application.
Knowing the 
details of the model, 
in particular the stencil types, at compile time allows the system to simplify expressions, 
and run common subexpression elimination to reduce the number of floating point operations (FLOPs) drastically.

\begin{figure}[htb]
	\centering
	\resizebox{\linewidth}{!}
	{
		\begin{tikzpicture}[framed,background rectangle/.style={fill=lssblue!10, rounded corners, draw=black!90, dashed}]

\definecolor{color1}{HTML}{3079A0}
\definecolor{color2}{HTML}{75B1B0}
\definecolor{color3}{HTML}{C5B353}
\definecolor{color4}{HTML}{D18855}
\def\scale{1.3};
\def\w{3};
\def\lw{.85mm};

\tikzset{
	state/.style={
		rectangle split ,
		rectangle split parts=2,
		rectangle split horizontal,
		rectangle split draw splits=false,
		rounded corners,
		draw=black, very thick,
		minimum height=4em,
		text width=2.55cm,
		inner sep=2pt,
		text centered,
	}
}

\tikzset{
	state2/.style={
		rectangle split ,
		rectangle split parts=3,
		rectangle split horizontal,
		rectangle split draw splits=false,
		rounded corners,
		draw=black, very thick,
		minimum height=4em,
		text width=1.64cm,
		inner sep=2pt,
		text centered,
	}
}

\tikzset{
	state3/.style={
		draw=black,very thick,
		rounded corners =.3cm,
		minimum height=4em,
		text width=2cm,
		text centered,
	}
}

\tikzset{
	state4/.style={
		rounded corners,
		draw=black, very thick,
		minimum height=4em,
		text width=5.2cm,
		text centered,
	}
}

	\node (a) [state4, left color=color1, right color=color1!40!white] at (0,10) {\textbf{lbmpy}};
	\node (b) [state3, fill=color1] at (-3.9, 10) {Model creation};
	
	\node (c) [state3, fill=color2, below=of b] {Code generation and optimisation};
	\node (d) [state2, left color=color2, right color=color2!40!white, below=of a] {\nodepart{one} Compute kernel \nodepart{two} Boundary conditions
	\nodepart{three} Communi-cation};
	
	\node (e) [state3, fill=color3, below=of c] {Backends};
	\node (f) [state4, left color=color3, right color=color3!40!white, below=of d] {CPU: LLVM and GCC GPU: CUDA and OpenCL};

	\node (g) [state3, fill=color4, below=of e] {Execution};
	\node (h) [state, left color=color4, right color=color4!40!white, below=of f] {\nodepart{one} Interactively with IPython \nodepart{two} MPI distributed with \textsc{waLBerla}};

	\draw[-stealth, line width= \lw, color1]  (0, 9.3) to[out=270, in=90]  (-2,8.3);
	\draw[-stealth, line width= \lw, color1]  (0, 9.3) to[out=270, in=90]  (0,8.3);
	\draw[-stealth, line width= \lw, color1]  (0, 9.3) to[out=270, in=90]  (2,8.3);

	\draw[-stealth, line width= \lw, color2]  (-2, 6.82) to[out=270, in=90]  (0,5.82);
	\draw[-stealth, line width= \lw, color2]  (0, 6.82) to[out=270, in=90]  (0,5.82);
	\draw[-stealth, line width= \lw, color2]  (2, 6.82) to[out=270, in=90]  (0,5.82);
	
	\draw[-stealth, line width= \lw, color3]  (0, 4.35) -- (-1,3.39);
	\draw[-stealth, line width= \lw, color3]  (0, 4.35) -- (1,3.39);

\end{tikzpicture}
	}
	\caption{Complete workflow of combining \lbmpy{} and \walberla{} for MPI parallel execution. Furthermore, \lbmpy{} can be used as a stand-alone package for prototyping.}
	\label{fig:lbmpywithwalberla}
\end{figure}
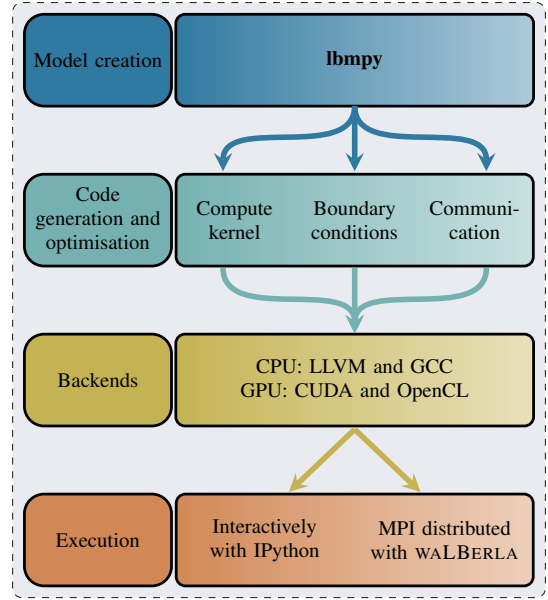

An overview of the complete workflow including the 
combination of \lbmpy{} and \walberla{} for MPI parallel execution is 
illustrated in \cref{fig:lbmpywithwalberla}. 
As described above the creation of the phase-field model is 
accomplished directly with \lbmpy{}, which forms a convenient prototyping environment 
since all equations can be stated as symbolic representations. 
\lbmpy{} does not only produce the compute kernels,  
but can generate also the pack and unpack information as it is needed for the MPI communication routines. 
This 
is again completely automatic, since the symbolic representations 
expose all field accesses and thus the 
data that must be kept in the ghost layers. A ghost layer is a single layer of cells around each subdomain used for the communication between neighboring subdomains.
Furthermore, also the routines are generated, to implement the boundary conditions.

The complete generation process can be 
configured to produce C-Code or code for GPUs with CUDA and alternatively OpenCL.
These kernels can be directly called as python functions to be run in an interactive environment or combined with the HPC framework \walberla{}.

To demonstrate the usage of \lbmpy{} we show 
how the update rule for the hydrodynamic distribution functions $g_i$ is realized.
Following \citet{Mitchel2018} \cref{eq:lb_flow}, should be formulated 
as
\begin{align}
\label{eq:improved_collision}
\begin{split}
\boldsymbol{g} (\boldsymbol{x} &+ \boldsymbol{e}_i \Delta t, t + \Delta t) = \boldsymbol{M}^{-1} \left[\boldsymbol{m} \right. \\
&- \left. \left(\boldsymbol{m}^{\mathrm{eq}} + 0.5 \boldsymbol{F}_{m}\right) \boldsymbol{S} + \boldsymbol{F}_{m}\right],
\end{split}
\end{align}
where 
	$\boldsymbol{m} = \boldsymbol{M}\boldsymbol{g}$
and the forcing is given by 
	$\boldsymbol{F}_m = \rho^{-1} \left(0, F_x, F_y, F_z, 0, \dots\right)$.
This formulation drastically reduces the number for FLOPs needed in each cell compared to \eqref{eq:lb_flow} where the force is applied in the particle distribution function (PDF) space. 
Note here that this kind of modification can be implemented in \lbmpy{} in a very simple way.
After creating the LB method it contains the moment matrix $\boldsymbol{M}$,
the relaxation matrix $\boldsymbol{S}$ and the moment equilibrium values $m_i^{\mathrm{eq}}$.
These variables are stated in \emph{SymPy} and can be used to directly write \cref{eq:improved_collision}.

\vspace{0.3cm} 
\begin{Verbatim}[commandchars=\\\{\},fontsize=\footnotesize, frame=single]
\PY{n}{method} \PY{o}{=} \PY{n}{lbmpy}\PY{o}{.}\PY{n}{create\PYZus{}lb\PYZus{}method}\PY{p}{(}\PY{p}{)}

\PY{n}{M} \PY{o}{=} \PY{n}{method}\PY{o}{.}\PY{n}{moment\PYZus{}matrix}
\PY{n}{S} \PY{o}{=} \PY{n}{method}\PY{o}{.}\PY{n}{relaxation\PYZus{}rates}
\PY{n}{m\PYZus{}eq} \PY{o}{=} \PY{n}{method}\PY{o}{.}\PY{n}{moment\PYZus{}equilibrium\PYZus{}values}
\PY{n}{F} \PY{o}{=} \PY{n}{hydrodynamic\PYZus{}force}\PY{p}{(}\PY{p}{)}
\PY{n}{m} \PY{o}{=} \PY{n}{M} \PY{o}{*} \PY{n}{g}

\PY{n}{g} \PY{o}{=} \PY{n}{m} \PY{o}{\PYZhy{}} \PY{p}{(}\PY{n}{m} \PY{o}{\PYZhy{}} \PY{n}{m\PYZus{}eq} \PY{o}{+} \PY{l+m+mf}{0.5} \PY{o}{*} \PY{n}{F}\PY{p}{)} \PY{o}{*} \PY{n}{S} \PY{o}{+} \PY{n}{F}
\PY{n}{g} \PY{o}{=} \PY{n}{g} \PY{o}{*} \PY{n}{M}\PY{o}{.}\PY{n}{inv}\PY{p}{(}\PY{p}{)}
\end{Verbatim}

\vspace{0.3cm} 

\subsection{Algorithm}
To discuss how 
the model of \cref{sec:model_description} can be realized, we will first 
present a straightforward implementation. 
The corresponding algorithm 
is displayed in \cref{alg:naiveAlgorithm} and will be discussed briefly in the following.
We start the time loop with time step size $\Delta t$, after initializing all fields. 
For MPI-parallel simulations, \walberla{} uses a domain partitioning into subdomains that are assigned to CPUs/GPUs \citep[see][]{waLBerla}. 
Thus, we perform the collision for both LB steps on each subdomain. 
Following that, we communicate the relevant PDF values of the ghost layers on each process for both PDF fields. 
Next, the streaming step for the phase-field LB step is 
executed.
In order to update the phase-field, we will then calculate the sum of the phase-field PDFs for each cell, 
according to \cref{eq:updatePhaseField}. 
Before we finalize the streaming step for the hydrodynamic LB step, we do the communication of the phase-field $\phi$. As the last step, we update the velocity field with the first-order moments of the hydrodynamic PDFs according to \cref{eq:updateVelocityField}.
To update the macroscopic variables $\phi$ and $\boldsymbol{u}$ each LB field has to be written one more time.

\IncMargin{.35em}
\begin{algorithm}[htb]
	Initialisation of all fields\;
	\For{each time step t}{
		Perform collision of phase-field PDFs\;
		Perform collision of velocity PDFs\;
		\BlankLine
		Communicate phase-field PDFs\;
		Communicate velocity PDFs\;
		\BlankLine
		Perform streaming of phase-field PDFs\;
		Update phase-field\;
		\BlankLine
		Communicate phase-field\;
		\BlankLine
		Perform streaming of velocity PDFs\;
		Update the velocity\;
	}
	\caption{Straightforward algorithm for the conservative ACM.}
	\label{alg:naiveAlgorithm}
\end{algorithm}
\DecMargin{.35em}

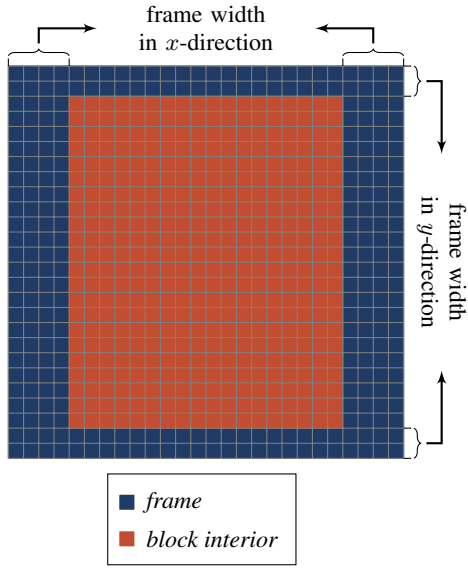
\begin{figure}[htb]
	\centering
	\begin{tikzpicture}[scale=0.02,
bluenode/.style={shape=rectangle, draw=white, fill=lssblue, line width=1},
rednode/.style={shape=rectangle, draw=white, fill=lssred, line width=1}]

\tikzset{
	myarrow/.style={->, >=latex', shorten >=1pt, thick},
}  

\pgfmathsetmacro{\OWx}{40}
\pgfmathsetmacro{\OWy}{20}
\draw[fill=lssblue, lssblue] (0,0) rectangle (260, 260);

\draw[fill=lssred, lssred] (\OWx,\OWy) rectangle (260-\OWx,260-\OWy);

\draw[step=10.0,black!50,thin,xshift=0cm,yshift=0cm] (0,0) grid (260,260);

\matrix [draw,below right] at (65, -10) {
	\node [bluenode,label=right:\emph{frame}] {}; \\
	\node [rednode,label=right:\emph{block interior}] {}; \\
};

\draw (260,0) -- (265,0);
\draw (260,20) -- (265,20);
\draw[decoration={brace,mirror,raise=5pt},decorate]
(258,0) -- (258,20);

\draw (260,240) -- (265,240);
\draw (260,260) -- (265,260);
\draw[decoration={brace,mirror,raise=5pt},decorate]
(258,240) -- (258,260);

\node [rotate=270, text width=3cm, align=center] at (285, 130) {frame width in $y$-direction};

\draw[myarrow] (275, 10) --  ++(10,0) -- (285, 60);	
\draw[myarrow] (275, 250) --  ++(10,0) -- (285, 200);

\draw (0,260) -- (0,265);
\draw (40,260) -- (40,265);
\draw[decoration={brace,raise=5pt},decorate]
(0,258) -- (40,258);

\draw (220,260) -- (220,265);
\draw (260,260) -- (260,265);
\draw[decoration={brace,raise=5pt},decorate]
(220,258) -- (260,258);

\node [rotate=0, text width=3cm, align=center] at (130, 285) {frame width in $x$-direction};

\draw[myarrow] (20, 275) --  ++(0,10) -- (60, 285);	
\draw[myarrow] (240, 275) --  ++(0,10) -- (200, 285);

\end{tikzpicture}
	\caption{Subdivision of the domain for communication hiding.}
	\label{fig:overlappingwidth}
\end{figure}

Based on this straightforward algorithm, we will now outline substantial improvements that can be made. 
To lower the memory footprint of the phase-field model, we combine the collision and
the streaming of the phase-field distribution functions and the update of the phase-field into one phase-field LB step.
Accordingly, the collision and the streaming of the velocity PDFs and
the update of the velocity field get combined to one hydrodynamic LB step. 
In this manner, the phase-field and velocity PDFs, as well as the phase-field and velocity field, get updated in only two instead of six compute kernels. 
A detailed overview of the proposed algorithm for the conservative ACM is presented in \cref{alg:conservativePhaseField}. 
In our proposed algorithm, we subdivide each LB step into iterations over an outer and an inner domain,
similarly to \citet{FEICHTINGER20151} but with variable cell width.
This is pointed out in \cref{fig:overlappingwidth}. 
For illustration purposes,
the figure 
illustrates only the two-dimensional case. 
The three-dimensional case is completely analogous. 
As shown, the frame width controls the iteration space of the outer and inner domain.
This can be done for all directions independently.
In the 
case illustrated, we have 
a frame width of four cells in $x$-direction and two grid cells in $y$-direction.

After the initialization of all fields, we start the time loop with time step size $\Delta t$.
Communication and computation can now be overlapped.
While updating the block interior of each subdomain with a stream-pull-collide scheme, 
we start the communication of the phase-field PDFs in the ghost layers.
Since the update for the phase-field is the zeroth-moment of the phase-field PDFs
as described by \cref{eq:updatePhaseField}, we also resolve the summation in the same kernel to minimize the memory footprint. 
Once the computation is completed,
we wait for the communication to finish and update the frame of each subdomain.\par

When the LB step for interface tracking is carried out,
we start the communication of the velocity-based distribution function together
with the phase-field. 
Note that these communication requirements can now be combined so that 
only one MPI-message must be sent.
Simultaneously, the inner part of the domain gets updated with the hydrodynamic LB step in a collide-stream-push manner. 
According to \cref{eq:viscousForce}, we need to form the non-equilibrium for the viscous force,
which makes a collide-stream-push scheme more convenient to use.
To lower the memory pressure, we update the velocity field with \cref{eq:updateVelocityField} in the same kernel accordingly. 
To finish a single time step, we wait for the communication and update the outer part of the domain.
Consequently, in \cref{alg:conservativePhaseField}, a one-step two-grid algorithm is applied for both LB steps. \par

\IncMargin{.35em}
\begin{algorithm}[htb]
	Initialisation of all fields\;
	\For{each time step t}{
		Start communication of phase-field PDFs\;
		Perform phase-field LB step inner domain\;
		\BlankLine
		Wait for the communication to finish\;
		Perform phase-field LB step outer domain\;
		\BlankLine
		\BlankLine
		Start communication of the velocity PDFs and the phase-field\;
		Perform hydrodynamic LB step inner domain\;
		\BlankLine
		Wait for the communication to finish\;
		Perform hydrodynamic LB step outer domain\;
	}
	\caption{Improved algorithm for the conservative ACM.}
	\label{alg:conservativePhaseField}
\end{algorithm}
\DecMargin{.35em}

\section{Benchmark Results}
\label{sec:benchmark}
In the following, we compare \cref{alg:naiveAlgorithm} and \cref{alg:conservativePhaseField}. For the straightforward algorithm, we need to load each PDF field three times during a single time step. In the improved algorithm, it is only necessary to load each field once. As we will discuss in \cref{sec:single_GPU} in more detail, the performance limiting factor of the model is the memory bandwidth. Therefore, we expect an increase in the performance of our improved algorithm approximately by a factor of three. To measure the performance of both algorithms, we initialize a squared domain of $260^3$ cells on an \nvidia{} Tesla P100 GPU. The domain consists of liquid with $\rho_H = 1$ and a gas bubble in the middle with $\rho_L = 0.001$ with a radius of $R = 65$ grid cells. The mobility is set to $M = 0.02$, the interface thickness is chosen to be $\xi = 5$, the surface tension is $\sigma = 10^{-4}$ and the relaxation time is $\tau = 0.53$. In all directions, periodic boundary conditions are applied, and there is no external force acting on the bubble. The same parameter setup is used for all other benchmarks in the next sections if not stated differently. We measure the performance in Mega Lattice Updates per Second (MLUPs) after $200$ time steps. For the unoptimized algorithm, we measure $211$~MLUPs, and for the improved algorithm, we measure $550$~MLUPs. Unlike expected, we can not entirely observe an improvement of a factor of three between the two algorithms. One reason might be that caching works better in the straight forward algorithm due to simpler compute kernels.\par

In the following sections we first investigate the performance of \cref{alg:conservativePhaseField} for a single GPU. Afterwards, we analyze the scaling behavior on an increasing number of GPUs in a weak scaling benchmark. For all investigations, we use a D3Q15 SRT LB scheme for the interface tracking and a D3Q27 MRT LB scheme for the velocity distribution function similarly to \citet{MITCHELL20181}.

\subsection{Single GPU}
\label{sec:single_GPU}

\begin{table*}[htb]
	\footnotesize
	\caption[Estimated performance results of the LB steps.]{Estimated performance results of the phase-field and hydrodynamic LB step. The memory bandwidth $b_\mathrm{s}$ is determined with a STREAM copy benchmark. The peak performance $p_{\mathrm{peak}}$ is given by the vendor \citep[see][]{V100}.}
	\label{tab:singleGPU}
	\centering
	\begin{tabular}{l l c c c c c c c c}
		\toprule
		Hardware  & LB kernel & \begin{tabular}[c]{@{}c@{}}$b_\mathrm{s}$ \\ \nicefrac{GB}{s}\end{tabular} & \begin{tabular}[c]{@{}c@{}}$p_{\mathrm{peak}}$\\ TFLOPS\end{tabular} & \begin{tabular}[c]{@{}c@{}}$n_\mathrm{b}$\\ bytes\end{tabular} & \begin{tabular}[c]{@{}c@{}}$n_\mathrm{f}$\\ FLOPS\end{tabular} & $B_\mathrm{c}$ & $B_\mathrm{m}$ & $l$   & \begin{tabular}[c]{@{}c@{}}Kernel estimate \\ GLUPs\end{tabular} \\ \midrule 
		\multirow{2}{*}{V100} & phase & 808 & 7.80 & 280 & 320  & 0.88 & 0.10 & 0.11 & 2.89 \\ \cmidrule{2-10} 
		& hydro & 808 & 7.80 & 488 & 809 & 0.60 & 0.10  & 0.16 & 1.66 \\ \midrule 
		\multirow{2}{*}{P100} & phase & 542 & 5.30 & 280 & 320 & 0.88 & 0.10 & 0.11 & 1.94 \\ \cmidrule{2-10} 
		& hydro & 542 & 5.30 & 488 & 809 & 0.60 & 0.10 & 0.16 & 1.11 \\ \bottomrule
	\end{tabular}%
\end{table*}

\begin{table*}[htb]
	\footnotesize
	\caption[Calculated effective bandwidth $b_\mathrm{eff}$.]{Calculated effective bandwidth $b_\mathrm{eff}$ for loads and stores in comparison with the measured bandwidth $b_\mathrm{m}$ by the nvprof~\textsuperscript{\ref{nvprof}}. Additionally, the measured performance of the LB steps is given and compared with the estimated results of \cref{tab:singleGPU}.}
	\label{tab:eff_bandwidth}
	\centering
	\begin{tabular}{l l c c c c c c}
		\toprule
		Hardware & LB kernel & \begin{tabular}[c]{@{}c@{}}$b_\mathrm{m, reads}$ \\ \nicefrac{GB}{s}\end{tabular} & \begin{tabular}[c]{@{}c@{}}$b_\mathrm{m, writes}$ \\ \nicefrac{GB}{s}\end{tabular} & \begin{tabular}[c]{@{}c@{}}$b_\mathrm{eff, reads}$ \\ \nicefrac{GB}{s}\end{tabular} & \begin{tabular}[c]{@{}c@{}}$b_\mathrm{eff, writes}$ \\ \nicefrac{GB}{s}\end{tabular} & \begin{tabular}[c]{@{}c@{}}Kernel measured \\ GLUPs\end{tabular} & \begin{tabular}[c]{@{}c@{}}Ratio \\ \%\end{tabular} \\ \hline
		\multirow{2}{*}{V100} & phase & 451 & 337 & 404 & 340 & 2.66 & 92 \\ \cmidrule{2-8} 
		& hydro & 372 & 341 & 337 & 326 & 1.36 & 82    \\ \midrule
		\multirow{2}{*}{P100} & phase & 292 & 219 & 258 & 217 & 1.70 & 87 \\ \cmidrule{2-8} 
		& hydro & 255 & 228 & 237 & 230 & 0.96 & 86\\ \bottomrule
	\end{tabular}%
\end{table*}

For the performance analysis on a single GPU, we focus on the \nvidia{} Tesla V100 due to its wide distribution and its usage in the top supercomputers Summit\footnote{\url{https://www.olcf.ornl.gov/summit/}} and Sierra\footnote{\url{https://computing.llnl.gov/computers/sierra}}. Further, we discuss the performance on a \nvidia{} Tesla P100 because it is used in the \pizdaint{}\footnote{\url{https://www.cscs.ch/computers/piz-daint/}} supercomputer, where we ran a weak scaling benchmark which is shown in \cref{sec:weak_scaling}. In this section, the two LB steps are analysed independently. In order to determine whether the LB steps are memory- or compute-bound, the balance model is used, which is based on the code balance $B_\mathrm{c}$
\begin{align}
B_\mathrm{c} &= \frac{n_\mathrm{b}}{n_\mathrm{f}},
\end{align}
and the machine balance $B_\mathrm{m}$ \citep[see][]{HagerAndWellein}
\begin{align}
B_\mathrm{m} &= \frac{b_\mathrm{s}}{p_{\mathrm{peak}}}.
\end{align}
The machine balance describes the ratio of the machine bandwidth $b_\mathrm{s}$ in bytes to the peak performance $p_{\mathrm{peak}}$ in FLOPs. The code balance, on the other hand, describes the ratio of bytes $n_\mathrm{b}$ loaded and stored for the execution of the algorithm to the executed FLOPs $n_\mathrm{f}$. The limiting factor of the algorithm is given by
\begin{align}
l = \min \left(1, \frac{B_\mathrm{m}}{B_\mathrm{c}}\right).
\end{align}
If the ``light speed'' balance $l$ is less than one a code is memory limited. To be able to calculate $l$, values for $b_\mathrm{s}$, $p_{\mathrm{peak}}$, $n_\mathrm{b}$ and $n_\mathrm{f}$ need to be stated.\par

As specified by the vendor, the V100 has a nominal bandwidth of $900$~\nicefrac{GB}{s} \citep[see][]{V100}. By running a STREAM copy benchmark, we get $808$~\nicefrac{GB}{s} as the stream copy bandwidth. This synthetic benchmark implements a vector copy $a_i = b_i$ and describes the behavior of the LB steps more realistically than the nominal bandwidth of the GPUs \citep[see][]{FEICHTINGER20151,bauer2020lbmpy}. Therefore, we will only refer to the stream copy bandwidth in further discussions. For the P100 a stream copy bandwidth of $542$~\nicefrac{GB}{s} can be measured in the same way.\par

The peak performances for the accelerator hardware is taken from the white paper by \nvidia{} \citep[see][]{V100}. For the V100 a double-precision peak performance of $7.8$~TFLOPs can be found while it is $5.3$~TFLOPs for the P100. \par

To determine $n_\mathrm{b}$ for the phase-field step, we first need to think about the data which needs to be read and written in a single cell in each iteration. In each cell, we update the phase-field PDFs in a stream-pull-collide manner. This means we need to read and write $15$~double-precision values per time step. Furthermore,  the velocity field is required in this calculation. Hence, another three double-precision values have to be loaded. Additionally, we need to take the forcing term into account as it is described by \cref{eq:interfaceTracking}. In this term, we approximate the curvature of the phase-field with a second order isotropic stencil as introduced in \cref{eq:grad,eq:lap}. This results in a 15-point stencil for the phase.field LB step. To estimate a lower limit for the memory traffic, we assume an ideal situation, which is reached when every grid point of the phase-field needs to be loaded only once. This would be the case if per cell only one value is loaded, and all additional values can be reused from cache since other threads already loaded them. Finally, we evaluate the zeroth-moment of the phase-field distributions to update the phase-field. Therefore, one more double-precision value is stored. Thus, we have in total $19$~double-precision values to load and 16 to store. Altogether this makes $280$~bytes per cell per iteration for the phase-field LB step.\par

For the hydrodynamic LB step, we have a D3Q27 stencil resulting in $27$~reads and $27$~writes. Further, we evaluate the velocity-field, and also update it. Thus another three loads and stores need to be performed.
Once again, we assume the ideal scenario of only one load when calculating the gradient and the Laplacian of \cref{eq:forceHydro}. This assumption leads to $31$ loads and $30$ stores. Hence, $488$ bytes per cell are needed per cell per iteration for the hydrodynamic LB step. \par

In order to get the number of operations, which are executed in one iteration per cell, $n_\mathrm{f}$ we use the \texttt{count\_ops}-function provided by \lbmpy{}. For the phase-field LB step, we get a total of $320$~FLOPs. Due to a larger stencil, a more complicated collision operator and a force model consisting of several terms, we get more operations for the hydrodynamic LB step namely $809$~FLOPs. These values are obtained after applying common subexpression eliminations. \par

\begin{figure}[htb]
	\centering
	\subfloat[Tesla V100]{
\begin{tikzpicture}[scale=1]
\tikzset{font=\small}
\definecolor{color0}{rgb}{0.121,0.231,0.4}
\definecolor{color1}{rgb}{0.76,0.3,0.19}

\begin{axis}[
height=6cm,
width=7cm,
legend cell align={left},
legend style={at={(0.03,0.97)}, anchor=north west, draw=white!80.0!black},
tick align=outside,
tick pos=left,
x grid style={white!69.01960784313725!black},
xmajorgrids,
xmin=-0.49, xmax=1.89,
xtick style={color=black},
xtick={0, 1.4},
xticklabels={Function unit (double),Memory (Device)},
y grid style={white!69.01960784313725!black},
ylabel={Utilisation},
ymajorgrids,
yticklabel={\pgfmathparse{\tick}\pgfmathprintnumber{\pgfmathresult}\%},
ymin=0, ymax=100,
ytick style={color=black}
]
\draw[fill=color0,draw opacity=0] (axis cs:0,0) rectangle (axis cs:-0.4,28);
\draw[fill=color0,draw opacity=0] (axis cs:1.4,0) rectangle (axis cs:1,97);
\draw[fill=color1,draw opacity=0] (axis cs:0,0) rectangle (axis cs:0.4,35);
\draw[fill=color1,draw opacity=0] (axis cs:1.4,0) rectangle (axis cs:1.8,88);
\node at (axis cs:-0.2,0)[
  scale=1.0,
  anchor=south,
  text=white,
  font=\bfseries,
  rotate=0.0
]{28\,\%};
\node at (axis cs:1.2,0)[
  scale=1.0,
  anchor=south,
  text=white,
  font=\bfseries,
  rotate=0.0
]{97\,\%};
\node at (axis cs:0.2,0)[
  scale=1.0,
  anchor=south,
  text=white,
  font=\bfseries,
  rotate=0.0
]{35\,\%};
\node at (axis cs:1.6,0)[
  scale=1.0,
  anchor=south,
  text=white,
  font=\bfseries,
  rotate=0.0
]{88\,\%};

\addlegendimage{ybar,ybar legend,fill=color0,draw opacity=0};
\addlegendentry{\scriptsize{phase-field LB step}}
\addlegendimage{ybar,ybar legend,fill=color1,draw opacity=0};
\addlegendentry{\scriptsize{hydrodynamic LB step}}
f\end{axis}

\end{tikzpicture}
		\label{fig:performanceV100} 
	}
	
	\subfloat[Tesla P100]{
\begin{tikzpicture}[scale=1]
\tikzset{font=\small}
\definecolor{color0}{rgb}{0.121,0.231,0.4}
\definecolor{color1}{rgb}{0.76,0.3,0.19}

\begin{axis}[
height=6cm,
width=7cm,
legend cell align={left},
legend style={at={(0.03,0.97)}, anchor=north west, draw=white!80.0!black},
tick align=outside,
tick pos=left,
x grid style={white!69.01960784313725!black},
xmajorgrids,
xmin=-0.49, xmax=1.89,
xtick style={color=black},
xtick={0, 1.4},
xticklabels={Function unit (double),Memory (Device)},
y grid style={white!69.01960784313725!black},
ylabel={Utilisation},
ymajorgrids,
yticklabel={\pgfmathparse{\tick}\pgfmathprintnumber{\pgfmathresult}\%},
ymin=0, ymax=100,
ytick style={color=black}
]
\draw[fill=color0,draw opacity=0] (axis cs:0,0) rectangle (axis cs:-0.4,30);
\draw[fill=color0,draw opacity=0] (axis cs:1.4,0) rectangle (axis cs:1,93);
\draw[fill=color1,draw opacity=0] (axis cs:0,0) rectangle (axis cs:0.4,33);
\draw[fill=color1,draw opacity=0] (axis cs:1.4,0) rectangle (axis cs:1.8,89);
\node at (axis cs:-0.2,0)[
  scale=1.0,
  anchor=south,
  text=white,
  font=\bfseries,
  rotate=0.0
]{30\,\%};
\node at (axis cs:1.2,0)[
  scale=1.0,
  anchor=south,
  text=white,
  font=\bfseries,
  rotate=0.0
]{93\,\%};
\node at (axis cs:0.2,0)[
  scale=1.0,
  anchor=south,
  text=white,
  font=\bfseries,
  rotate=0.0
]{33\,\%};
\node at (axis cs:1.6,0)[
  scale=1.0,
  anchor=south,
  text=white,
  font=\bfseries,
  rotate=0.0
]{89\,\%};

\addlegendimage{ybar,ybar legend,fill=color0,draw opacity=0};
\addlegendentry{\scriptsize{phase-field LB step}}
\addlegendimage{ybar,ybar legend,fill=color1,draw opacity=0};
\addlegendentry{\scriptsize{hydrodynamic LB step}}
\end{axis}

\end{tikzpicture}
		\label{fig:performanceP100}
	}
	\caption[Performance limiter measured with nvprof.]{Compute units utilization and memory transfer measured with nvprof \textsuperscript{\ref{nvprof}}. The memory transfer is based on the STREAM copy bandwidth of the hardware. The measurements are conducted for the phase-field and the hydrodynamic LB step seperatly on a Tesla V100 (a) and a Tesla P100 (b).}
	\label{fig:limiting_factors}
\end{figure}
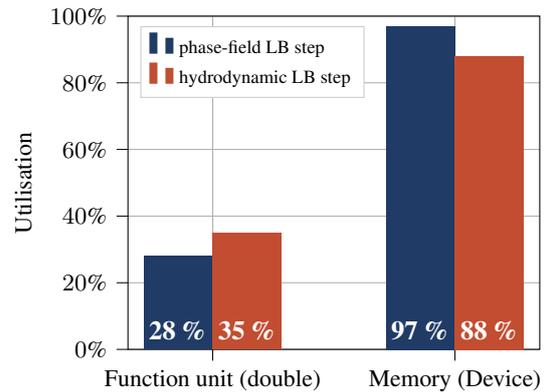
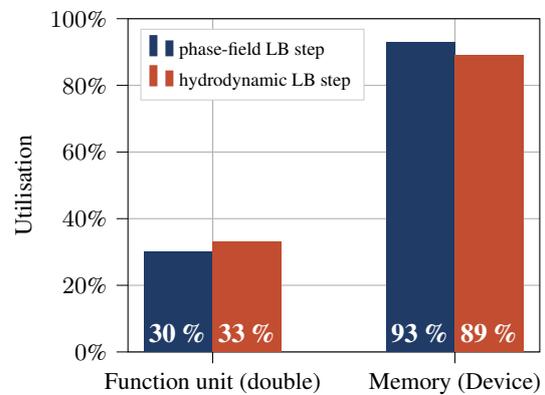

Combining the obtained values in \cref{tab:singleGPU}, we can see that both LB steps are highly memory bound. Therefore, the maximal performance is given by
\begin{align}
\label{eq:roofline}
P_{\mathrm{max}} = \frac{b_\mathrm{s}}{n_\mathrm{b}}.
\end{align} 
Additionally, we profiled the two LB steps with the \nvidia{} profiler nvprof\footnote{\label{nvprof}\url{https://docs.nvidia.com/cuda/profiler-users-guide/index.html}} for both GPU architectures. The results for the limiting factors are illustrated in \cref{fig:limiting_factors}. These measurements confirm the theoretical performance model very well, and show that the memory bandwidth is almost fully utilized for both GPUs. 

Consequently, the next step is to determine if the memory bandwidth is also reasonable. This means no unnecessary values should be transferred if they are not needed for the calculation. By running the phase-field LB step independently, we measure a performance of about $2.65$~GLUPs. With this information, we can calculate the effective bandwidth $b_\mathrm{eff}$ for reads and writes by multiplying the measured performance with the transferred data, respectively. For reads, this results in $404$~\nicefrac{GB}{s}, while for writes it results in $340$~\nicefrac{GB}{s}. This matches well with the results measured by the \nvidia{} profiler, which are $451$~\nicefrac{GB}{s} and $337$~\nicefrac{GB}{s}, respectively. It shows that there is not much unnecessary bandwidth utilization. Still, the measured bandwidth for loading values from the device memory is slightly higher than the calculated effective bandwidth. This indicates that the ideal assumption of every grid point of the phase-field being read only once is not completely true. One reason might be due to the high memory transfer caused by the underlying LB step. However, comparing the measured performance to the estimate in \cref{tab:singleGPU}, we reach about $92$~\% of the theoretical peak performance, which means that there is not much potential for further improvements. For the hydrodynamic LB step and for the P100 GPU, the results of the measured bandwidth and the effective bandwidth are gathered in \cref{tab:eff_bandwidth}. As one can see, the behavior of both LB steps is similar on both architectures. Moreover, both LB steps reach around $85$~\% of the theoretical peak performance on both GPUs.

\begin{figure}[htb]
	\centering
	\subfloat[Performance on a Tesla V100]{

\begin{tikzpicture}[scale=1]
\tikzset{font=\small}
\definecolor{color0}{rgb}{0.121,0.231,0.4}

\begin{axis}[
height=6cm,
width=7cm,
tick align=outside,
tick pos=left,
x grid style={white!69.01960784313725!black},
xmajorgrids,
xmin=-0.59, xmax=3.59,
xtick style={color=black},
xtick={0,1,2,3},
xticklabels={{(27, 27)},{(19, 19)},{(15, 15)},{(9, 9)}},
y grid style={white!69.01960784313725!black},
ylabel={MLUPs},
ymajorgrids,
ymin=0, ymax=2356.2,
ytick style={color=black}
]
\draw[fill=color0,draw opacity=0] (axis cs:-0.4,0) rectangle (axis cs:0.4,730);
\draw[fill=color0,draw opacity=0] (axis cs:0.6,0) rectangle (axis cs:1.4,1012);
\draw[fill=color0,draw opacity=0] (axis cs:1.6,0) rectangle (axis cs:2.4,1223);
\draw[fill=color0,draw opacity=0] (axis cs:2.6,0) rectangle (axis cs:3.4,2147);
\path [draw=black, very thick]
(axis cs:-0.4,842)
--(axis cs:0.4,842);

\path [draw=black, very thick]
(axis cs:0.6,1148)
--(axis cs:1.4,1148);

\path [draw=black, very thick]
(axis cs:1.6,1403)
--(axis cs:2.4,1403);

\path [draw=black, very thick]
(axis cs:2.6,2244)
--(axis cs:3.4,2244);

\node at (axis cs:0,0)[
  scale=1.0,
  anchor=south,
  text=white,
  font=\bfseries,
  rotate=0.0
]{86\,\%};
\node at (axis cs:1,0)[
  scale=1.0,
  anchor=south,
  text=white,
  font=\bfseries,
  rotate=0.0
]{88\,\%};
\node at (axis cs:2,0)[
  scale=1.0,
  anchor=south,
  text=white,
  font=\bfseries,
  rotate=0.0
]{87\,\%};
\node at (axis cs:3,0)[
  scale=1.0,
  anchor=south,
  text=white,
  font=\bfseries,
  rotate=0.0
]{95\,\%};
\end{axis}

\end{tikzpicture}
		\label{fig:performancev100} 
	}
	
	\subfloat[Performance on a Tesla P100]{
\begin{tikzpicture}[scale=1]
\tikzset{font=\small}
\definecolor{color0}{rgb}{0.121,0.231,0.4}

\begin{axis}[
height=6cm,
width=7cm,
tick align=outside,
tick pos=left,
x grid style={white!69.01960784313725!black},
xmajorgrids,
xmin=-0.59, xmax=3.59,
xtick style={color=black},
xtick={0,1,2,3},
xticklabels={{(27, 27)},{(19, 19)},{(15, 15)},{(9, 9)}},
y grid style={white!69.01960784313725!black},
ylabel={MLUPs},
ymajorgrids,
ymin=0, ymax=1581.3,
ytick style={color=black}
]
\draw[fill=color0,draw opacity=0] (axis cs:-0.4,0) rectangle (axis cs:0.4,491);
\draw[fill=color0,draw opacity=0] (axis cs:0.6,0) rectangle (axis cs:1.4,672);
\draw[fill=color0,draw opacity=0] (axis cs:1.6,0) rectangle (axis cs:2.4,809);
\draw[fill=color0,draw opacity=0] (axis cs:2.6,0) rectangle (axis cs:3.4,1410);
\path [draw=black, very thick]
(axis cs:-0.4,565)
--(axis cs:0.4,565);

\path [draw=black, very thick]
(axis cs:0.6,770)
--(axis cs:1.4,770);

\path [draw=black, very thick]
(axis cs:1.6,940)
--(axis cs:2.4,940);

\path [draw=black, very thick]
(axis cs:2.6,1506)
--(axis cs:3.4,1506);

\node at (axis cs:0,0)[
  scale=1.0,
  anchor=south,
  text=white,
  font=\bfseries,
  rotate=0.0
]{86\,\%};
\node at (axis cs:1,0)[
  scale=1.0,
  anchor=south,
  text=white,
  font=\bfseries,
  rotate=0.0
]{87\,\%};
\node at (axis cs:2,0)[
  scale=1.0,
  anchor=south,
  text=white,
  font=\bfseries,
  rotate=0.0
]{86\,\%};
\node at (axis cs:3,0)[
  scale=1.0,
  anchor=south,
  text=white,
  font=\bfseries,
  rotate=0.0
]{93\,\%};
\end{axis}

\end{tikzpicture}
		\label{fig:performancep100}
	}
	\caption[Performance measurement for different LB stencils.]{Performance measurement for different LB stencils (D3Q27, D3Q19, D3Q15 and D2Q9) for the phase-field and the hydrodynamic LB step compared to the theoretical peak performances which are illustrated as black lines respectively. The white number in each bar shows the ratio between theoretical peak performance and measured performance.}
	\label{fig:performance}
\end{figure}
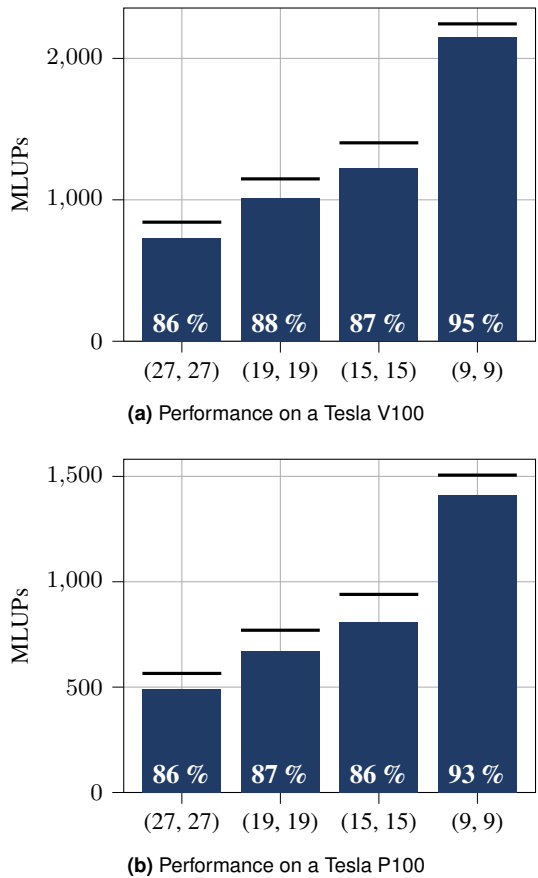

With the usage of code generation, we can easily change the discrete velocities used for the LB steps of the implementation. This makes it possible for us to evaluate the performance for different two and three-dimensional stencils. As shown in \cref{fig:performance} we employed a D3Q27, D3Q19, D3Q15, and a D2Q9 stencil for both the phase-field and the hydrodynamic LB step. We show that we are able to reach about $86$~\% of the theoretical peak performance for different three-dimensional stencils. This number increases to about $94$~\% on a Tesla V100 and on a Tesla P100 in the two-dimensional case because the assumption that every value of the phase-field $\phi$ is loaded only once is better fulfilled in those cases. The ideal assumption that each value of the phase-field $\phi$ is only loaded once is better fulfilled in the two-dimensional case. Thus, we reach a higher relative performance of about $94$~\% on a Tesla V100 and on a Tesla P100.

\subsection{Weak Scaling Benchmark}
\label{sec:weak_scaling}

\begin{figure*}[htb]
	\centering
\begin{tikzpicture}

\definecolor{color0}{rgb}{0.121,0.231,0.4}
\definecolor{color1}{rgb}{0.76,0.3,0.19}
\definecolor{color2}{rgb}{0.22,0.44,0.57}

\begin{axis}[
height=7cm,
legend cell align={left},
legend style={at={(0.03,0.03)}, anchor=south west, draw=white!80.0!black},
log basis x={2},
tick align=outside,
tick pos=left,
width=14cm,
xlabel={GPUs},
xmajorgrids,
xmin=0.683020128377198, xmax=2998.44750529664,
xmode=log,
xtick style={color=black},
ylabel={MLUPs per GPU},
ymajorgrids,
ymin=0, ymax=750,
ytick style={color=black}
]
\addplot [very thick, color0, mark=*, mark size=3, mark options={solid}]
table {%
1 507.23229678478
2 504.773980913487
4 502.183237625483
8 496.455528973479
16 495.15647269082
32 494.811136107269
64 496.10208728687
128 498.19206846662
256 496.235602346269
512 497.190171723918
1024 473.236327775565
2048 496.514022282794
};
\addlegendentry{frame width $(32, 8, 8)$}
\addplot [very thick, color1, mark=triangle*, mark size=3, mark options={solid}, dashed]
table {%
1 441.822999745613
2 439.979660883784
4 436.968135989162
8 436.385255156783
16 436.524303606838
32 435.602604672788
64 435.700228850306
128 434.806977367555
256 435.543419813185
512 435.083898117253
1024 435.221676922235
2048 435.758818831836	
};
\addlegendentry{frame width $(1, 1, 1)$}
\addplot [very thick, color2, mark=diamond*, mark size=3, mark options={solid}, dotted]
table {%
1 549.435555375762
2 503.866261922388
4 448.742400342177
8 381.457295909547
16 390.855196721032
32 383.587342066074
64 358.472648716384
128 376.941287089822
256 371.964461855022
512 354.517889492298
1024 345.996009657575
2048 348.972000388536
};
\addlegendentry{no communication hiding}
\addplot [very thick, gray, mark options={solid}]
table {%
	1 705
	2 705
	4 705
	8 705
	16 705
	32 705
	64 705
	128 705
	256 705
	512 705
	1024 705
	2048 705
};
\addlegendentry{theoretical maximum}
\end{axis}

\end{tikzpicture}
	\caption{Weak scaling performance benchmark on the \pizdaint{} supercomputer. The grey line shows the theoretical peak performance. With a thicker frame width (dark blue) we reach a parallel efficiency of almost $98$~\% and also $70$~\% of the theoretical peak performance. Furthermore, it can be seen, that a thin frame (red) shows worse performance and no separation of the domain (light blue) shows worse parallel efficiency.}
	\label{fig:weakscaling}
\end{figure*}
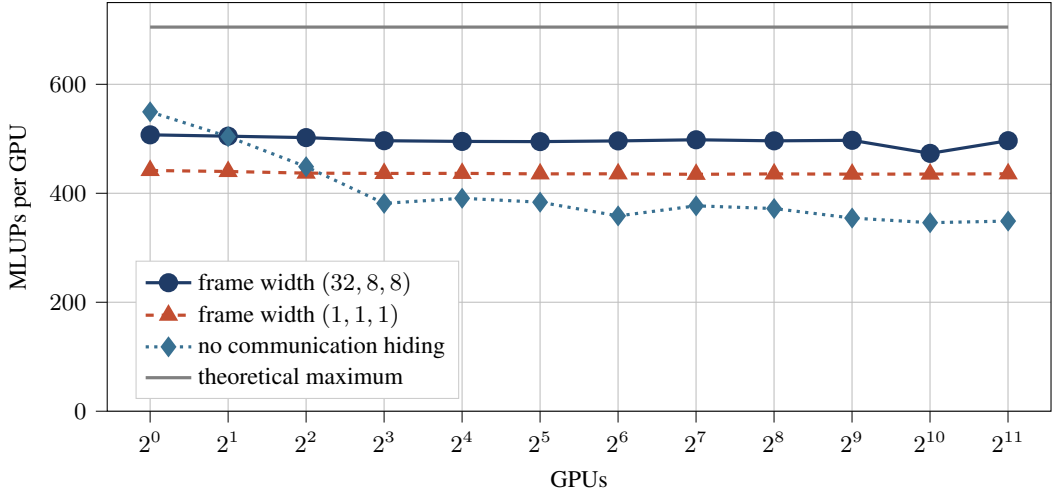

The idea of communication hiding for LBM simulations 
was already studied in \citet{FEICHTINGER20151}.
As in \citet{FEICHTINGER20151} we partition each subdomain (block) into an outer part,
to obtain a \emph{frame} for each block, and an inner part, the \emph{block interior}. 
In comparison to that, it is possible in our work, to choose the width of the block frame freely. We can then execute the LBM kernels first on the frame, then send it asynchronously as ghost values to the neighbouring blocks. While the communication takes place, the LBM kernels can be executed on the inner domains. Due to our flexible implementation, we ran benchmarks for different frame widths on an increasing number of GPUs in order to find an optimal choice of the frame width. This weak scaling performance benchmark was carried out on the \pizdaint{} supercomputer on up to $2048$~GPU nodes. We set the physical parameters and the number of grid cells on each GPU as discussed at the beginning of this section. Thus we always have $260^3$ grid cells on each GPU. This benchmark is performed with different frame widths to determine their performance. As we increase the number of nodes, we get more communication overhead. It follows that a frame width of` $(32, 8, 8)$ shows a significantly higher performance than an algorithm without communication hiding as pictured in \cref{fig:weakscaling}. On a single GPU, on the other hand, we can see that the performance of the algorithm without communication hiding is higher with 550~GLUPs. Configurations with a thicker frame width like $(64, 64, 64)$ would perform better on a single GPU than a frame width of $(32, 8, 8)$ but show no good scaling behavior similar to when we do not utilize communication hiding at all. Thus for the sake of simplicity, we only show the scaling behavior of three scenarios in \cref{fig:weakscaling}. A frame width of $(32, 8, 8)$, as it delivers a good scaling behavior and performance. The native case describes a frame width of $(1, 1, 1)$ and ``no communication hiding'' refers to a test case where we do not utilize communication hiding strategies at all. Furthermore, the theoretical peak performance, which is calculated without taking the communication overhead into account, is shown as a grey line in the figure.

Both, a frame width of $(32, 8, 8)$ and a frame width of $(1, 1, 1)$ show excellent scaling behavior. However, since the kernel to update the outer frame for the native case is not as efficient as in the other case, we see lower performance throughout the benchmark. If we do not utilize communication hiding at all, it is visible that we lose performance when scaling up our simulation. On $2048$~GPUs, we run a simulation setup with about $36$~billion lattice cells and one TLUPs. Furthermore, we reach a parallel efficiency of about $98$~\%. We are still able to exploit $70$~\% of the theoretical peak performance on $2048$ nodes. In comparison, we would be able only to achieve $62$~\% of the maximum performance when using a frame width of $(1, 1, 1)$. Not using communication hiding reduces the measured performance to $49$~\% of the theoretical peak performance. This lets us conclude that we can save a significant amount of compute-time and resources by using our flexible implementation, which is not only possible for one specific LB configuration but for a wide variety of different setups to solve the LBEs including state of the art collision operators. On the other side, the user has a very convenient interactive development environment to work on new problems and can almost entirely work directly on the equation level.

\section{Numerical Validation}
\label{sec:numericalResults}

\subsection{Single Rising Bubble}
\label{sec:single_rising_bubble}
The motion of a single gas bubble rising in liquid has been studied for many centuries by various authors and is still a problem of great interest today \citep[see][]{Clif78, bhaga_weber_1981, Grace73, Tomiyama98, Lote18, FAKHARI_weighted_LBM, Mitchel19}. This is due to its vast importance in many industrial applications and natural phenomena like the aerosol transfer from the sea, oxygen dissolution in lakes due to rain, bubble column reactors and the flow of foams and suspension, to name just a few. Because of the three-dimensional nature and nonlinear effects of the problem, the numerical simulation still remains a challenging task \citep[see][]{Tripathi}. The evolution of the gas bubble in stationary fluids depends on a large variety of different parameters. These are the surface tension, the density difference between the fluids, the viscosity of the bulk media and the external pressure gradient or gravitational field through which buoyancy effects are observed in the gas phase. The parameters are developed into dimensionless groups in order to acquire comprehensive theories describing the problem \citep[see][]{Mitchel19}. \par

\begin{figure}[htb]
	\subfloat[$\mathrm{Eo} = 1$, $\mathrm{Re}_{\mathrm{Gr}} = 40$]{
		\includegraphics[width=0.45\linewidth]{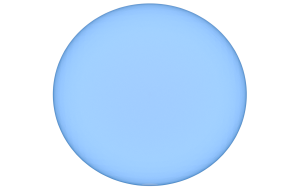}
		\label{fig:a}
	}
	\subfloat[$\mathrm{Eo} = 30$, $\mathrm{Re}_{\mathrm{Gr}} = 10$]{
		\includegraphics[width=0.45\linewidth]{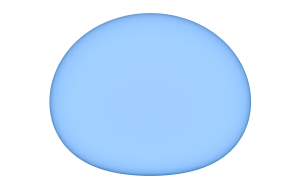}
		\label{fig:b}
	}
	
	\subfloat[$\mathrm{Eo} = 5$, $\mathrm{Re}_{\mathrm{Gr}} = 40$]{
		\includegraphics[width=0.45\linewidth]{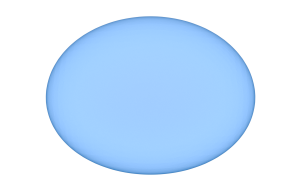}
		\label{fig:c}
	}
	\subfloat[$\mathrm{Eo} = 30$, $\mathrm{Re}_{\mathrm{Gr}} = 30$]{
		\includegraphics[width=0.45\linewidth]{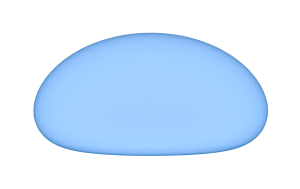}
		\label{fig:d}
	}
	
	\subfloat[$\mathrm{Eo} = 100$, $\mathrm{Re}_{\mathrm{Gr}} = 40$]{
		\includegraphics[width=0.45\linewidth]{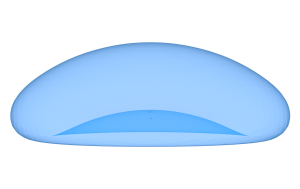}
		\label{fig:e}
	}
	\subfloat[$\mathrm{Eo} = 30$, $\mathrm{Re}_{\mathrm{Gr}} = 120$]{
		\includegraphics[width=0.45\linewidth]{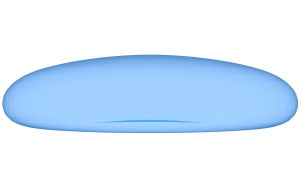}
		\label{fig:f}
	}
	
	\caption{Terminal shape of a single rising bubble with $\rho^* = 1000$ and $\mu^* = 100$ under different Eötvös and gravity Reynolds numbers at $10 t^*$} 
	\label{fig:bubble_shapes}
\end{figure}

In this study we set up a computational domain of $256 \times 1024 \times 256$ cells and initially place a spherical bubble with a radius of $16$ grid cells at $(128, 256, 128)$. We use no-slip boundary conditions at the top and bottom of our domain to mimic solid walls. In all other directions, periodic boundary conditions are applied. This setup is consistent with \citet{Mitchel19}. For the force acting on the bubble we use the volumetric buoyancy force
\begin{align}
\boldsymbol{F}_\mathrm{b} = \rho g_y \boldsymbol{\hat{y}},
\end{align} 
where $g_y$ is the magnitude of the gravitational acceleration, which is applied in the vertical direction $\boldsymbol{\hat{y}}$ \citep[see][]{FAKHARI_weighted_LBM}. The density $\rho$ is calculated by a linear interpolation with \cref{eq:density}. For all simulations, a D3Q19 stencil is used for both LB steps with an MRT method and a weighted orthogonal moment set \citep[see][]{FAKHARI_weighted_LBM}. To characterise the shape of the bubble, we need five dimensionless parameters. We use the Reynolds number based on the gravitational force
\begin{align}
\mathrm{Re}_\mathrm{Gr} = \frac{\rho_H \sqrt{g_y D^3}}{\mu_H},
\end{align} 
where $D$ is the initial diameter of the bubble. The Eötvös number
\begin{align}
\mathrm{Eo} = \frac{g_y \rho_H D^2}{\sigma},
\end{align}
which is also called the Bond number describes the influence of gravitational forces compared to surface tension forces. Further, we use the density ratio $\rho^* = 1000$, the viscosity ratio $\mu^* = 100$, and the reference time 
\begin{align}
t_{\mathrm{ref}} = \sqrt{\frac{D}{g_y}}.
\end{align}
Thus, the dimensionless time can be calculated as $t^* = \nicefrac{t}{t_{\mathrm{ref}}}$ For different Eötvös and gravitational Reynolds numbers, the terminal shape of a rising bubble can be seen in \cref{fig:bubble_shapes}. All simulations are carried out with a reference time of $t_{\mathrm{ref}} = 18\,000$ until $t^* = 10$. We set the interface thickness to $\xi = 5$ cells and the mobility to $M = 0.04$. Comparing our results with \citet{Mitchel19}, we see a good agreement regarding the terminal shape of the bubble. Further, we achieve the same behavior as described by the experiments of \citet{bhaga_weber_1981}. The shape of the bubbles represents the increase of the effective force acting on them. As a result, we can observe a deformation from less spherical to flatter bubbles when increasing the Eötvös or the gravitational Reynolds number.\par

\begin{figure}[htb]
	\centering
	\input{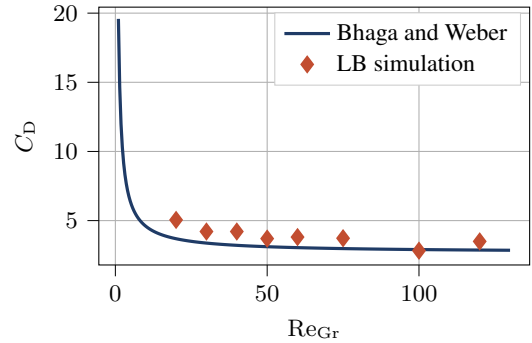}
	\caption[Drag coefficient of a single rising bubble.]{Drag coefficient plotted against the Reynolds number of a single rising bubble. The dots show the results of the LB simulation with a constant Eötvös number of $\mathrm{Eo} = 30$ while the blue line represents the analytical estimation from \cref{eq:analytical_drag_coeff} made by \citet{bhaga_weber_1981} in 1981.}
	\label{fig:dragcoefficient}
\end{figure}

Additionally, we calculate the drag coefficients with the terminal velocity of the bubbles
\begin{align}
C_\mathrm{D}^{\mathrm{LBM}} = \frac{4}{3} \frac{g_y \left(\rho_H - \rho_L\right) D}{\rho_H \, u_t^2},
\end{align}
and compare it to the experiments carried out by \citet{bhaga_weber_1981}. Based on their observations they set up the following empirical equation to calculate the drag coefficient of a rising bubble described by the gravity Reynolds number
\begin{align}
\label{eq:analytical_drag_coeff}
C_\mathrm{D}^{\mathrm{exp}} = \left[2.67^{\frac{9}{10}} + \left(\frac{16}{\mathrm{Re}_{\mathrm{Gr}}}\right)^{\frac{9}{10}}\right]^{\frac{10}{9}}.
\end{align}
As it can be seen in \cref{fig:dragcoefficient} our results are in good agreement with the experimental investigations. 

\subsection{Bubble Field}
\label{sec:bubble_field}
\begin{figure*}[htb]
	\centering
	\subfloat[Initialisation]{
		\includegraphics[width=.49\linewidth]{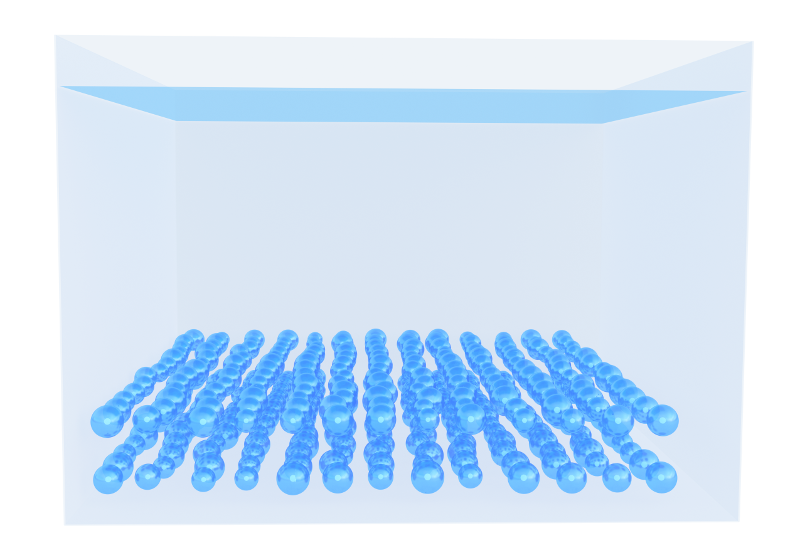}
		\label{fig:bubblefield0}
	}
	
	\subfloat[Time step \num{125 000}]{
		\includegraphics[width=.49\linewidth]{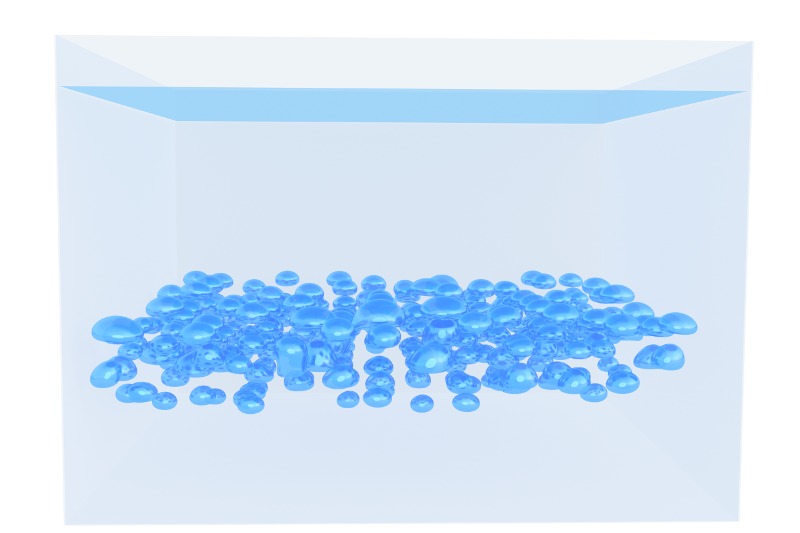}
		\label{fig:bubblefield1}
	}
	\subfloat[Time step \num{250 000}]{
		\includegraphics[width=.49\linewidth]{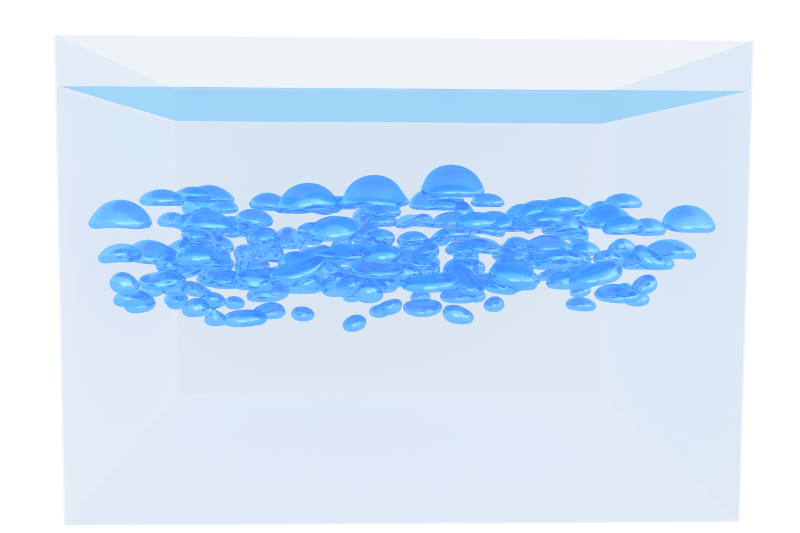}
		\label{fig:bubblefield2}
	}
	
	\subfloat[Time step \num{375 000}]{
		\includegraphics[width=.49\linewidth]{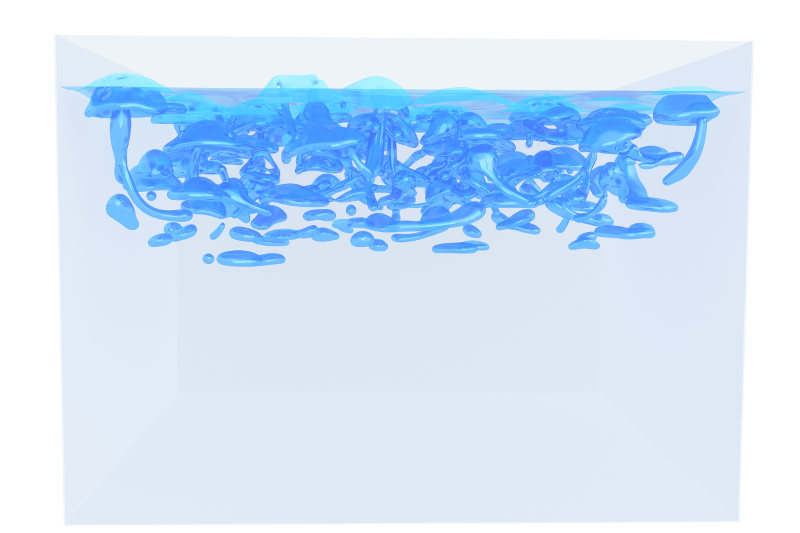}
		\label{fig:bubblefield3}
	}
	\subfloat[Time step \num{500 000}]{
		\includegraphics[width=.49\linewidth]{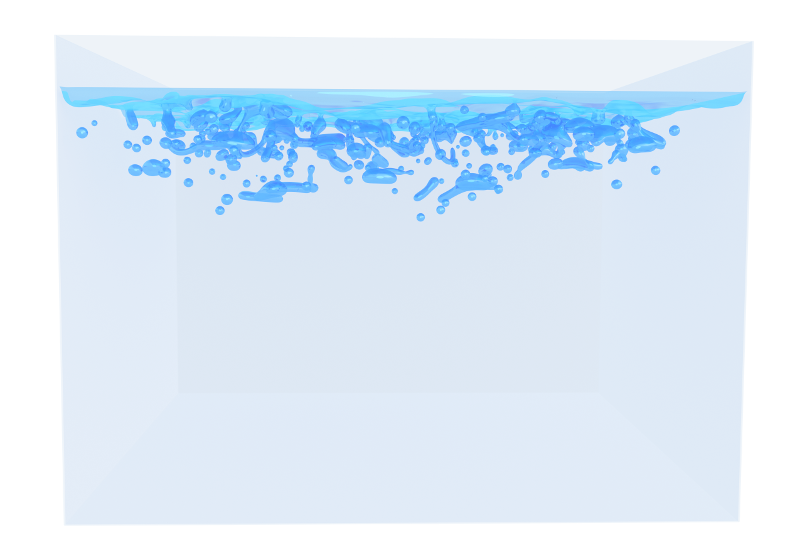}
		\label{fig:bubblefield}
	}
	\caption{Large scale bubble rise scenario simulated on the \pizdaint{} supercomputer with several hundred air bubbles.}
	\label{fig:test}
\end{figure*}

For demonstrating the robustness as well as our possibilities through the efficient and scalable implementation, we show a large scale bubble rise scenario with several hundred bubbles. The simulation is carried out on a $720 \times 560 \times 720$ domain, which gives about $290$ million lattice cells. Our simulation has run for $10$ hours resulting in \num{500000} time steps. For this simulation, we use an Eötvös number of $\mathrm{Eo} = 50$. We further specify the gravitational Reynolds number as $\mathrm{Re}_\mathrm{Gr} = 50$ and the mobility as $M = 0.08$. We set the reference time to $t^* = 18\,000$. The density ratio and the viscosity ratio between the fluid and the bubbles are set to the ones of water and air ($\rho^* = 1000$, $\mu^* = 100$). We initialize two layers of air bubbles at the bottom of our domain with a radius of $R = 16$. To initialize the bubbles with a slightly different radius, we add a random value sampled from $[-\nicefrac{R}{5}, \nicefrac{R}{5}]$ to the radius. By having air bubbles with different radii, we can see that bubbles with a larger radius accelerate faster, which shows a good physical agreement. A screenshot of the simulation every \num{125000} time steps can be seen in \cref{fig:bubblefield}. We can see clearly that complex physical phenomena are carried out in stable simulation. These are bubble coalescence, the coalescence of air bubbles with the liquid surface and bubble breakage.

\section{Conclusion}
\label{sec:Conclusion}
In this work we have shown an implementation of the conservative ACM based on meta-programming. With this technique we can generate highly efficient C, OpenCL and CUDA kernels which can be integrated in other frameworks for simulating large scale scenarios. For this work we have used the \textsc{waLBerla} framework to integrate our code. 
We have measured the efficiency our implementation on single GPUs. 
Excellent performance results compared to the roofline model could be shown 
for a Tesla V100 and a Tesla P100 where we could achieve about $85$ \% of 
the theoretical peak performance for both architectures.
Additionally, we have shown that our code not only performs very well for one configuration but keeps its excellent efficiency even for different stencils and different methods to solve the LBEs. It is even possible to directly generate 2D cases for testing which also show very good performance results. 
Through separating our iteration region in an inner and an outer part we could enable communication hiding, relevant for multi-GPU simulations with MPI. With this technique we are able to run large scale simulations with almost perfect scalability. 
To show this we have run a weak scaling benchmark on the \pizdaint{} supercomputer on $2048$ GPU nodes. We could show that our implementation has a parallel efficiency of almost $98$~\%.
To validate our code form a physical point of view we have measured the terminal shape
of single rising air bubble in water under various Eötvös numbers and Reynolds numbers.
We could show 
good agreement with the literature data regarding the terminal shape 
and the drag coefficient of rising bubbles. 
Finally, we have set up a larger scenario where we simulate several hundred air bubbles in the water. With our implementation, we were not only able to maintain a stable simulation for the complicated test case, but we could also observe complex phenomena like bubble coalescence, the coalescence of air bubbles with the liquid surface and bubble breakage. 

\section{Supporting information}

The multiphysics framework waLBerla is released as an open-source project and can be used under the terms of an GNU general public license. The source code is available at \url{https://i10git.cs.fau.de/walberla/walberla}.

\section{Acknowledgments}

We appreciate the support by Travis Mitchell for this project. Furthermore, we thank Christoph Schwarzmeier and Christoph Rettinger for fruitful discussions on the topic.

\section{Funding}

We are grateful to the Swiss National Supercomputing Centre (CSCS) for providing computational resources and access to the \pizdaint{} supercomputer. Further, the authors would like to thank the Bavarian Competence Network for Technical and Scientific High Performance Computing (KONWIHR), the Deutsche Forschungsgemeinschaft (DFG,
German Research Foundation) for supporting project 408059952 and the Bundesministerium für Bildung und Forschung (BMBF, Federal Ministry of Education and Research) for supporting project 01IH15003A (SKAMPY).

\bibliographystyle{SageH}
\bibliography{references.bib}

\end{document}